\documentclass[hidelinks,a4paper,10pt,twocolumn, switch]{article}

\usepackage{fancyhdr}
\fancypagestyle{plain}{%
  \fancyhf{}
\fancyfoot[R]{\thepage}
\fancyfoot[L]{\small \textit{Preprint submitted to Springer}\\ \texttt{arXiv}}
}

\usepackage[affil-it]{authblk}
\makeatletter
\def\@maketitle{%
  \newpage
  \null
  \vskip 2em%
  \begin{center}%
  \let \footnote \thanks
    {\Large\bfseries \@title \par}%
    \vskip 1.5em%
    {\normalsize
      \lineskip .5em%
      \begin{tabular}[t]{c}%
        \@author
      \end{tabular}\par}%
    \vskip 1em%
    {\normalsize \@date}%
  \end{center}%
  \par
  \vskip 1.5em}
\makeatother

\usepackage[a4paper, total={17.5cm,24cm}]{geometry}

\usepackage[font=normal,labelfont=bf]{caption}

\usepackage{sectsty} %small fonts for appendix by means of \sectionfont{\small}

\usepackage{titlesec}
\titlespacing\section{0pt}{12pt plus 3pt minus 3pt}{1pt plus 1pt minus 1pt}
\titlespacing\subsection{0pt}{10pt plus 3pt minus 3pt}{1pt plus 1pt minus 1pt}
\titlespacing\subsubsection{0pt}{8pt plus 3pt minus 3pt}{1pt plus 1pt minus 1pt}

\titleformat{\section}{\normalfont\large\bfseries}{\thesection}{1em}{}

\titleformat{\subsection}{\normalfont\normalsize\bfseries}{\thesubsection}{1em}{}

\titleformat{\subsubsection}{\normalfont\normalsize}{\thesubsubsection}{1em}{}

\titleformat{\paragraph}[runin]{\normalfont\normalsize\itshape}{\theparagraph}{1em}{}

\usepackage{newtxtext,newtxmath} % use Times fonts if available on your TeX system
\usepackage{mathtools,upgreek}

\usepackage[utf8]{inputenc}	% allow utf-8 input
\usepackage[T1]{fontenc}	% use 8-bit T1 fonts
\usepackage{xcolor}		% colors for hyperlinks
\usepackage[colorlinks = false]{hyperref}	% Color links to references, figures, etc.
\usepackage{booktabs} 		% professional-quality tables
\usepackage{nicefrac}		% compact symbols for 1/2, etc.
\usepackage{microtype}		% microtypography
\usepackage{lineno}		%   Line numbers
\usepackage{float}			% Allows for figures within multicol

\usepackage{siunitx}[=v2]
\usepackage{nicefrac}
\usepackage{subfig}
\usepackage{paralist}
\usepackage{algorithmic,algorithm2e}

\newcommand\numberthis{\addtocounter{equation}{1}\tag{\theequation}}
\mathtoolsset{centercolon}

\newcommand{\ea}{et al.}

\DeclareMathOperator{\tr}{tr}
\DeclareMathOperator{\dev}{dev}
\DeclareMathOperator{\sym}{sym}

\newcommand{\inte}[3]{\int \limits_{ #1} #2 \; \mathrm{d} #3} 
\newcommand{\binte}[4]{\mathop{\int}_{ #1}^{#2} #3 \; \mathrm{d} #4}

\newcommand{\ddp}{\, \partial} 

\newcommand{\nabr}{\nabla_{\ve{X}}}

\newcommand{\diffp}[2]{\frac{\partial #1}{\partial #2}}
\newcommand{\diffv}[2]{\frac{\delta #1}{\delta #2}}

\newcommand{\ve}[1]{\boldsymbol{#1}} 
\newcommand{\te}[1]{\mathbf #1}
\newcommand{\tg}[1]{\boldsymbol{#1}} 
\newcommand{\tte}[1]{\mathbb{#1}}

\newcommand{\tei}{\mathbf I}

\newcommand{\rset}{\mathbb{R}}
\newcommand{\nset}{\mathbb{N}}

\newcommand{\muss}{\stackrel{!}{=}}
\newcommand{\const}{\text{const}.}
\newcommand{\nv}{\mathrm}
\newcommand{\comma}{\hspace{3mm} \text{,}}

\newcommand{\glmand}{\hspace{3mm} \text{and} \hspace{3mm}}

\newcommand{\commam}{\hspace{3mm} \text{,} \hspace{3mm}}
\newcommand{\point}{\hspace{3mm} \text{.}}

\newcommand{\omref}{\varOmega_0}

\newcommand{\gc}{\mathcal{G}_\text{c}}
\newcommand{\gco}{\mathcal{G}_\text{c}^1}
\newcommand{\gct}{\mathcal{G}_\text{c}^2}
\newcommand{\rref}{r_\text{ref}}

\newcommand{\lc}{{\ell_\text{c}}}
\newcommand{\lcs}{{\ell^2_\text{c}}}
\newcommand{\etaf}{{\eta_\text{f}}}
\newcommand{\hist}{\mathcal{H}}
\newcommand{\xirr}{\ve X_\nv{irrBC}}

\newcommand{\bvi}{\beta_\text{vi}}

\newcommand{\gstd}{g_\text{st}(d)}
\newcommand{\gvid}{g_\text{vi}(d)}
\newcommand{\gst}{g_\text{st}}
\newcommand{\gvi}{g_\text{vi}}

\newcommand{\Pist}{\varPi^\nv{sd}}
\newcommand{\Pif}{\varPi^\nv{fr}}

\newcommand{\Pilc}{\varPi_\lc}
\newcommand{\Pistlc}{\varPi^\nv{sd}_\lc}
\newcommand{\Piflc}{\varPi^\nv{fr}_\lc}

\newcommand{\Phif}{\varPhi^\nv{fr}}
\newcommand{\Phifd}{\dot \varPhi^\nv{fr}}

\newcommand{\disua}{{D}^\nv{vi,1D}}

\newcommand{\disd}{\dot{D}}
\newcommand{\dpfd}{\dot{D}^\nv{fr}}
\newcommand{\dvid}{\dot{D}^\nv{vi}}

\newcommand{\psist}{\psi^\text{st}}
\newcommand{\psivi}{\psi^\text{vi}}

\newcommand{\psiste}{\psi^\text{st,eq}}
\newcommand{\psisto}{\psi^\text{st,ov}}

\newcommand{\psistev}{{}^\nv{vol}\psi^\text{st,eq}}
\newcommand{\psistei}{{}^\nv{iso}\psi^\text{st,eq}}

\newcommand{\psistov}{{}^\nv{vol}\psi^\text{st,ov}}
\newcommand{\psistoi}{{}^\nv{iso}\psi^\text{st,ov}}

\newcommand{\Psist}{\varPsi^\text{st}}

\newcommand{\Psivi}{\varPsi^\text{vi}}

\newcommand{\keq}{\kappa^\nv{eq}}
\newcommand{\kov}{\kappa^\nv{ov}}

\newcommand{\Neq}{N_\nv{O}^\nv{eq}}
\newcommand{\Nov}{N_\nv{O}^\nv{ov}}

\newcommand{\mueq}[1][p]{\mu^\nv{eq}_{#1}}
\newcommand{\muov}[1][p]{\mu^\nv{ov}_{#1}}
\newcommand{\mueqn}{\mu^\nv{eq}}
\newcommand{\muovn}{\mu^\nv{ov}}
\newcommand{\aleq}[1][p]{\alpha^\nv{eq}_{#1}}
\newcommand{\alov}[1][p]{\alpha^\nv{ov}_{#1}}

\newcommand{\nueqn}{\nu^\nv{eq}}
\newcommand{\nuovn}{\nu^\nv{ov}}

\newcommand{\visc}{\tte V}

\newcommand{\etav}{\prescript{\nv{vol}}{}{\eta}}
\newcommand{\etai}{\prescript{\nv{iso}}{}{\eta}}

\newcommand{\fiso}{\overline{\te F}}

\newcommand{\fel}{{\te{F}^\nv{el}}}
\newcommand{\jel}{{J^\nv{el}}}

\newcommand{\feli}{{\overline{\te F}^\nv{el}}}

\newcommand{\fvi}{{\te{F}^\nv{vi}}}
\newcommand{\fvid}{{\dot{\te{F}}^\nv{vi}}}

\newcommand{\bel}{{\te b}^\nv{el}}
\newcommand{\beltr}{{\te b}^\nv{el}_\nv{tr}}
\newcommand{\cel}{{\tilde{\te C}}^\nv{el}}

\newcommand{\cvi}{{{\te C}^\nv{vi}}}

\newcommand{\lame}[1][\beta]{{\lambda_{#1}^\nv{el}}}
\newcommand{\lamed}[1][\beta]{{\bar{\lambda}_{#1}^\nv{el}}}
\newcommand{\lamei}[1][\gamma]{{\bar \lambda^\nv{el}_{#1}}}

\newcommand{\lameq}[1][\beta]{{\lambda_{#1}^\nv{el}}^2}
\newcommand{\nlame}{{N^\nv{el}_\lambda}}
\newcommand{\nulame}[1][\gamma]{{\nu}^\nv{el}_{#1}}

\newcommand{\epse}[1][\beta]{{\varepsilon_{#1}^\nv{el}}}
\newcommand{\epsetr}[1][\beta]{{\varepsilon_{\nv{tr} \, {#1}}^\nv{el}}}
\newcommand{\epsetrdr}{{\varepsilon_{\nv{tr} \, 3}^\nv{el}}}

\newcommand{\pel}[1][\beta]{{\te p_{#1}^\nv{el}}}
\newcommand{\Pel}[1][\beta]{{\tilde{\te P}_{#1}^\nv{el}}}

\newcommand{\dvie}{{\te d}^\nv{vi}}
\newcommand{\lvi}{{\te l}^\nv{vi}}
\newcommand{\lvit}{{\tilde{\te{{l}}}{}^\nv{vi}}}
\newcommand{\dvit}{{\tilde {\te d}^\nv{vi}}}

\newcommand{\lie}{\mathcal{L}}

\newcommand{\ute}{\prescript{0}{}{\te T}^\nv{eq}}
\newcommand{\uto}{\prescript{0}{}{\te T}^\nv{ov}}

\newcommand{\utao}{\prescript{0}{}{\tg \uptau}^\nv{ov}}
\newcommand{\utaoj}[1][\beta]{\prescript{0}{}{\uptau}^\nv{ov}_{#1}}
\newcommand{\utaodev}[1][\beta]{\prescript{0}{}{\uptau}^\nv{ov,dev}_{#1}}

\newcommand{\wu}{\delta \ve u}
\newcommand{\wuj}[1][j]{\delta u_{#1}}

\newcommand{\wusj}{\mathbb{W}_{u_j}}

\newcommand{\wc}{\delta c}

\newcommand{\wcs}{\mathbb{W}_c}

\newcommand{\domrefuj}{\partial \varOmega_{0 \, u_j}}

\newcommand{\coeq}{\prescript{0}{}{\tte{C}}^\nv{eq}}
\newcommand{\coov}{\prescript{0}{}{\tte{C}}^\nv{ov}}

\newcommand{\fetr}{{{\te F}^\nv{el}_\nv{tr}}}
\newcommand{\fvio}{\prescript{}{n-1}{\te{F}^\nv{vi}}}
\newcommand{\fvioab}[1]{\prescript{}{n-1}{{F}^\nv{vi}_{#1}}}
\newcommand{\celtrh}{{\breve{\te C}}^\nv{el}_\nv{tr}}
\newcommand{\celtrhab}[1]{{\breve{C}}^\nv{el}_\nv{tr #1}}

\newcommand{\utoh}{\prescript{0}{}{\breve{\te T}}^\nv{ov}}
\newcommand{\utohab}[1]{\prescript{0}{}{\breve{T}}^\nv{ov}_{#1}}

\newcommand{\coovh}{\prescript{0}{}{\breve{\tte{C}}}^\nv{ov}}

\newcommand{\Neh}[1][\alpha]{\breve{\ve N}_{#1}}
\newcommand{\lametr}[1][\alpha]{{\lambda_{\nv{tr} \,#1}^\nv{el}}}

\title{Phase-field modelling and analysis of rate-dependent fracture phenomena at finite deformation}

\begin{document}

\author{Franz Dammaß, \quad Karl A. Kalina, \quad Marreddy Ambati, \quad Markus Kästner
  \thanks{Contact: \texttt{markus.kaestner@tu-dresden.de} }}
\affil{Institute of Solid Mechanics, \\ TU Dresden, Germany}

\date{}
\maketitle

\begin{abstract}
Fracture of materials with rate-dependent mechanical behaviour, e.g. polymers, is a highly complex process. For an adequate modelling, the coupling between rate-de\-pen\-dent stiffness, dissipative mechanisms present in the bulk material and crack driving force has to be accounted for in an appropriate manner.
In addition, the fracture toughness, i.e. the resistance against crack propagation, can depend on rate of deformation.

In this contribution, an energetic phase-field model of rate-dependent fracture at finite deformation is presented. For the deformation of the bulk material, a formulation of finite viscoelasticity is adopted with strain energy densities of Ogden type assumed.
The unified formulation allows to study different expressions for the fracture driving force. Furthermore, a possibly rate-dependent toughness is incorporated.
The model is calibrated using experimental results from the literature for an elastomer and predictions are qualitatively and quantitatively validated against experimental data. Predictive capabilities of the model are studied for monotonic loads as well as creep fracture.
Symmetrical and asymmetrical crack patterns are discussed and the influence of a dissipative fracture driving force contribution is analysed. It is shown that, different from ductile fracture of metals, such a driving force is not required for an adequate simulation of experimentally observable crack paths and is not favourable for the description of failure in viscoelastic rubbery polymers.
Furthermore, the influence of a rate-dependent toughness is discussed by means of a numerical study. From a phenomenological point of view, it is demonstrated that rate-dependency of resistance against crack propagation can be an essential ingredient for the model when specific effects such as rate-dependent brittle-to-ductile transitions shall be described.

\vspace{5mm}
\noindent
\textbf{Keywords: } Phase-field~\textendash~Fracture~\textendash~Elastomers~\textendash~Rate-dependent fracture toughness \textendash~Viscoelasticity~\textendash~Dissipation~\textendash~Finite deformation

\end{abstract}

\section{Introduction}
\label{sec:intro}
The mechanical behaviour of many engineering materials depends on rate of deformation. For example, the response of polymers can be much more stiff or brittle when the loading rate is increased, see \cite{grellmann2015,gent2012}.
The same applies for natural materials such as cheese \cite{goh2005} or confections \cite{vanvliet2005}.
In order to reduce experimental effort for design and testing of engineering products as well as for the optimisation of production processes of foods, the computational modelling and simulation of crack phenomena in rate-dependent materials is of increasing interest.

For the modelling of crack phenomena, the phase-field approach to fracture has become a well-established concept.
Different from classical finite element approaches (FE), it enables to simulate crack growth without the need for remeshing. Furthermore, complex crack patterns which are not a priori known can be simulated in a straightforward manner, which especially makes the concept attractive compared to alternative approaches such as Cohesive zone elements \cite{ortiz1999} or the Extended-finite-element-method \textit{(X-FEM)} \cite{moes2017}.
The phase-field fracture approach goes back to the variational formulation of brittle fracture of Francfort and Marigo~\cite{francfort1998}, who recast the Griffith criterion \cite{griffith1921} for crack propagation into a variational setting.
Bourdin~\ea~\cite{bourdin2000,bourdin2008} introduced a diffuse crack representation by means of the phase-field variable, which continuously varies from the intact to the fully broken material state. In other words, cracks are no longer seen as sharp discontinuities, but approximated over a finite length scale $\lc$. Making use of this smeared crack representation, a regularisation of the pseudo-energy functional is carried out. 
Based upon the fundamental work of Bourdin, numerous phase-field models of brittle fracture have been proposed which include several advancements within both the infinitesimal strain regime \cite{miehe2010a,miehe2010,kuhn2010,amor2009,steinke2019} as well as finite deformation \cite{weinberg2017,mang2021a,swamynathan2022}.
Further extensions have been proposed to also include fatigue effects, see \cite{alessi2018,carrara2020,seiler2020,schreiber2020}, inter alia.
Very recently, the variational approach to fracture also is combined with machine learning and data-driven approaches \cite{goswami2020a,feng2021,aldakheel2021,carrara2020b,carrara2021}. 

Furthermore, fracture phase-field modelling has been advanced towards elasto-plastic materials, see~\cite{alessi2018a} for an overview on several approaches within the infinitesimal strain setting. 
For the performance of ductile fracture models, the description of interaction between inelastic dissipative mechanisms and crack growth has revealed crucial. In particular, in the absence of an adequate coupling, crack patterns that are experimentally observed in metals, for instance, can not be reproduced, see e.g. \cite[Fig.~14]{ambati2015}.
Different manners of introducing such a coupling are proposed, including 
non-energetic ductile fracture driving forces based on accumulated plastic strain \cite{miehe2015,schaenzel2015}, 
and an enhanced degradation function which, in addition to the phase-field variable, depends on plastic deformation and results in a distinct plastic contribution to the fracture driving force \cite{ambati2015,ambati2016b}.
Furthermore, instead of a fracture driving force related to inelastic mechanisms, degradation of fracture toughness depending on equivalent plastic strain is introduced \cite{yin2020c}.
Several other phase-field models of ductile fracture are based upon a pseudo-energy functional in which both elastically stored energy and a plastic quantity, which is referred to as plastic work or plastic energy, are assumed to degrade upon fracture.
Depending on the specific formulation, the plastic contribution to free energy actually corresponds to hardening terms~\cite{kuhn2016,miehe2016a} or accumulated plastic deformation~\cite{borden2016}.

More recently, the approach is combined with rate-de\-pendent models for the deformation of the bulk material.
A first phase-field fracture model for viscoelastic solids is proposed by Schänzel~\cite{schaenzel2015}, where a non-energetic fracture driving force based on a generalised principal stress criterion is adopted.
Alternative driving forces based on energetic or thermodynamic arguments are introduced by Shen~\ea \cite{shen2019} as well as Liu~\ea \cite{liu2018} within the kinematically linear regime and by Loew~\ea \cite{loew2019,loew2020} within the linear viscoelasticity framework \cite{holzapfel1996a} at finite deformation. In these models, a viscous dissipative contribution is incorporated into the degraded free energy and thus enters fracture driving force.
Different from the aforementioned models, only equilibrium and over-stress parts of the strain energy density are assumed to promote crack propagation by Yin and Kaliske~\cite{yin2020b}, who combined the phase-field approach to fracture with a model of finite viscoelasticity~\cite{reese1998}.
Recently, similar formulations are adopted by Brighenti~\ea~\cite{brighenti2021} based on statistical mechanics-based equations for the response of the bulk material, and in \cite{arash2021} where the rate- and temperature-dependent behaviour of polymer nanocomposites is investigated.
In some of these models based on either non-energetic or energetically motivated driving forces, \cite{schaenzel2015,loew2019}, the viscosity assumed for the evolution of phase-field which originally is solely numerically motivated, cf. \cite{miehe2010,kuhn2010}, is understood as a material parameter and identified from experimental data.
In the recent work of Dammaß et al. \cite{dammass2021a,dammass2021b}, a unified energetic phase-field model for fracture of viscoelastic solids has been presented in the kinematically linear regime. Depending on the specific choice of the degradation functions and model parameters, respectively, the modelling approaches of \cite{shen2019}, \cite{loew2019,loew2020} or \cite{yin2020b,brighenti2021} are retained as limiting cases of the present model and by means of representative numerical studies, the coupling between viscous effects and fracture is analysed.

Compared to the rate-dependent behaviour of the bulk material, less efforts have been devoted to the study of strain rate-dependent resistance against fracture.
Miehe~\ea~\cite{miehe2015} suggested a phenomenological ansatz for the rate-dependent toughness in order to investigate the brittle-to-ductile fracture mode transition observed in the \textit{Kalthoff-Winkler} experiment, i.e. for shear-loaded metals.
Yin~\ea~\cite{yin2020g} assumed the toughness of a linear elastic material to depend on rate of deformation. In their formulation, dissipation due to crack formation is incorporated into the free energy so that additional stress contributions are obtained from the rate-dependent fracture toughness.
In these two models \cite{miehe2015,yin2020g}, rate-independent models for the deformation of the bulk are considered.
To the best of the authors' knowledge, so far, there are no phase-field models that consider both a rate-dependent toughness and a model of rate-dependent deformation.

\begin{figure*}[tb!]
		\centering 
		\includegraphics[trim=0mm 0 20mm 0,width=0.93\textwidth]{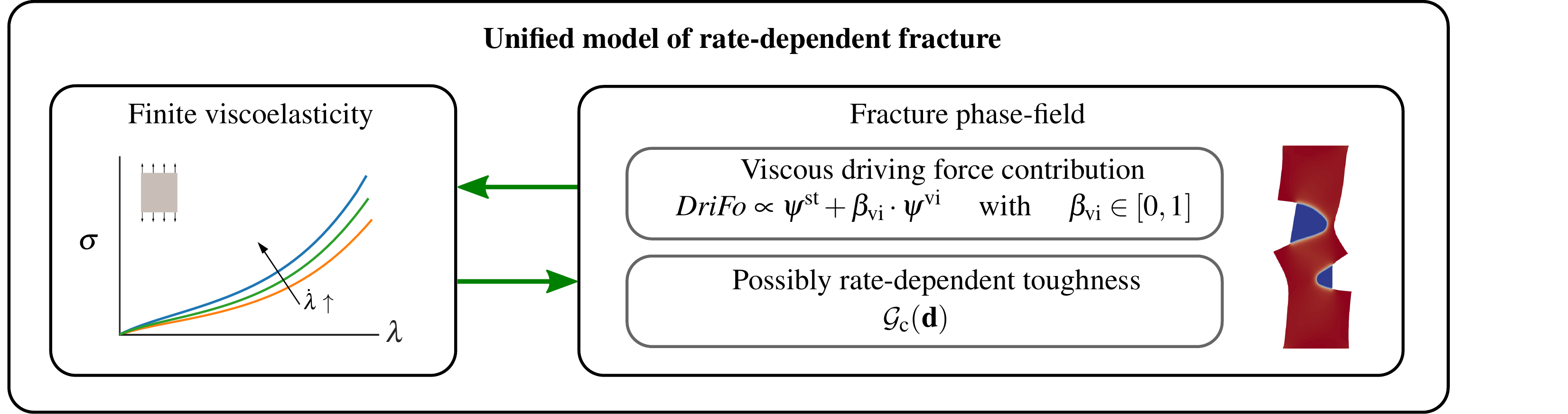} % [trim=left bottom right top]
		\caption{Modular structure and flexibility of the proposed model for rate-dependent fracture phenomena}
		\label{fig:model_overview}
\end{figure*}
In the present contribution, a thermodynamically consistent phase-field model for fracture of materials with rate-dependent behaviour is presented.
For this purpose, the previously introduced pseudo-energy functional \cite{dammass2021b}, which consists of the free energy that includes a contribution related to viscous dissipative mechanisms, and the fracture contribution is advanced towards the finite viscoelasticity setting of Reese and Govindjee \cite{reese1998}.
Depending on the specific choice of the model parameters, the modelling approaches of \cite{shen2019}, \cite{loew2019,loew2020} or \cite{yin2020b,brighenti2021} can be retained as limiting cases.
Based on experimental data for an Ethylene Propylene Diene Monomer (EPDM) rubber from the literature \cite{loew2019}, the model parameters for the response of the bulk and the fracture behaviour are identified and model predictions are qualitatively and quantitatively verified on experimental results.
In doing so, two assumptions for the fracture driving force, i.e. whether there shall be a contribution related to viscous dissipation or not, are investigated.
With the aim of studying the possible influence of such a driving force component on the crack path, an asymmetrical setup is studied in addition to the symmetrical ones considered in recent publications, e.g. \cite{loew2019,yin2020b}.
Furthermore, based on experimental evidence on strain rate-dependent fracture toughness, cf. \cite{goh2005,gamonpilas2009,forte2015}, and motivated from a phenomenological point of view, a rate-dependent fracture toughness is introduced.
A numerical study on the coupling between rate-dependent resistance against crack propagation and viscoelastic bulk response is then performed.
An overview on the structure of the proposed unified 
model is given in Fig.~\ref{fig:model_overview}.

% -- structure --
The paper proceeds as follows. In Sect.~\ref{sec:deriv}, the proposed phase-field model of fracture in rate-dependent materials at finite deformation is presented and its thermodynamic consistency is proven.
Subsequently, in Sect.~\ref{sec:impl}, algorithmic aspects are addressed.
In Sect.~\ref{sec:numres}, the model parameterization is described and various numerical examples serve for validation and analysis of the model.
A short summary and an outlook regarding the future work is given in Sect.~\ref{sec:concl}.
In the Appendix, information on the tangent for the \textit{local} Newton iteration and the \textit{global} material tangent is given.

% Notation
Within this paper, italic symbols are used for scalar quantities ($d$, $\varPsi$) and bold italic symbols for vectors ($\ve u$). For Second-order tensors, bold non-italic letters ($\te T$, $\tg \uptau$) are used, whereas fourth-order tensors are written in \textit{Blackboard bold}~($\tte C$).

\section{Phase-field formulation}
\label{sec:deriv}

In this Section, the phase-field model of fracture in materials with rate-dependent behaviour is presented.
At first, the general energetic formulation of fracture in viscoelastic materials derived in \cite{dammass2021b} is extended to the finite deformation setting.
Subsequently, the specific constitutive assumptions are outlined. %, including rate-dependent fracture toughness. % and thermodynamic consistency is proven.
Finally, governing equations are provided and thermodynamic consistency is proven.

\subsection{Pseudo-energy functional}
\label{sec:enfu}

\paragraph{The variational approach to fracture.}
Following the pioneering work of Griffith~\cite{griffith1921}, the dissipation due to crack growth $\Pif$ can be understood as an energetic quantity, which increases proportional to the crack surface.
Accordingly, a pseudo-energy functional%
\footnote{For sake of brevity, terms arising from external loads are omitted in~\eqref{eq:fun-nireg} and what follows.}
\begin{equation}
 \varPi = \inte{\omref  \backslash \varGamma_0}{\psi}{V} + \inte{\varGamma_0}{\gc}{A} \, =: \Pist + \Pif 
 \label{eq:fun-nireg}
\end{equation}
can be defined \cite{francfort1998}, wherein $\omref \subset \rset^N$ is the reference or undeformed configuration of the $N$-dimensional domain under consideration and \mbox{$\varGamma_0 \subset \omref$} denotes the corresponding crack surface.
The stored free energy is given by $\Pist$ and its density with respect to the reference configuration is denoted by $\psi$.
The proportionality coefficient $\gc>0$ typically is referred to as fracture toughness. While $\gc$ is assumed to be a constant in the classical theory, in the recent literature, it is assumed to change during fatigue life, see e.g. \cite{seiler2020}, or due to plastic deformation \cite{yin2020c,han2021}. Furthermore, it can explicitly depend on the position in space in heterogeneous materials~\cite{hansen-doerr2019a,hansen-doerr2020}.
In the following, the variational phase-field framework is set up for the case that $\gc$ is constant, first. Subsequently, the model is extended to account for a fracture toughness that depends on rate of deformation in Sect.~\ref{sec:gc_def}.

\begin{figure}[t!]
		\centering 
		\includegraphics[scale=0.6,trim=10mm 0 5mm 0]{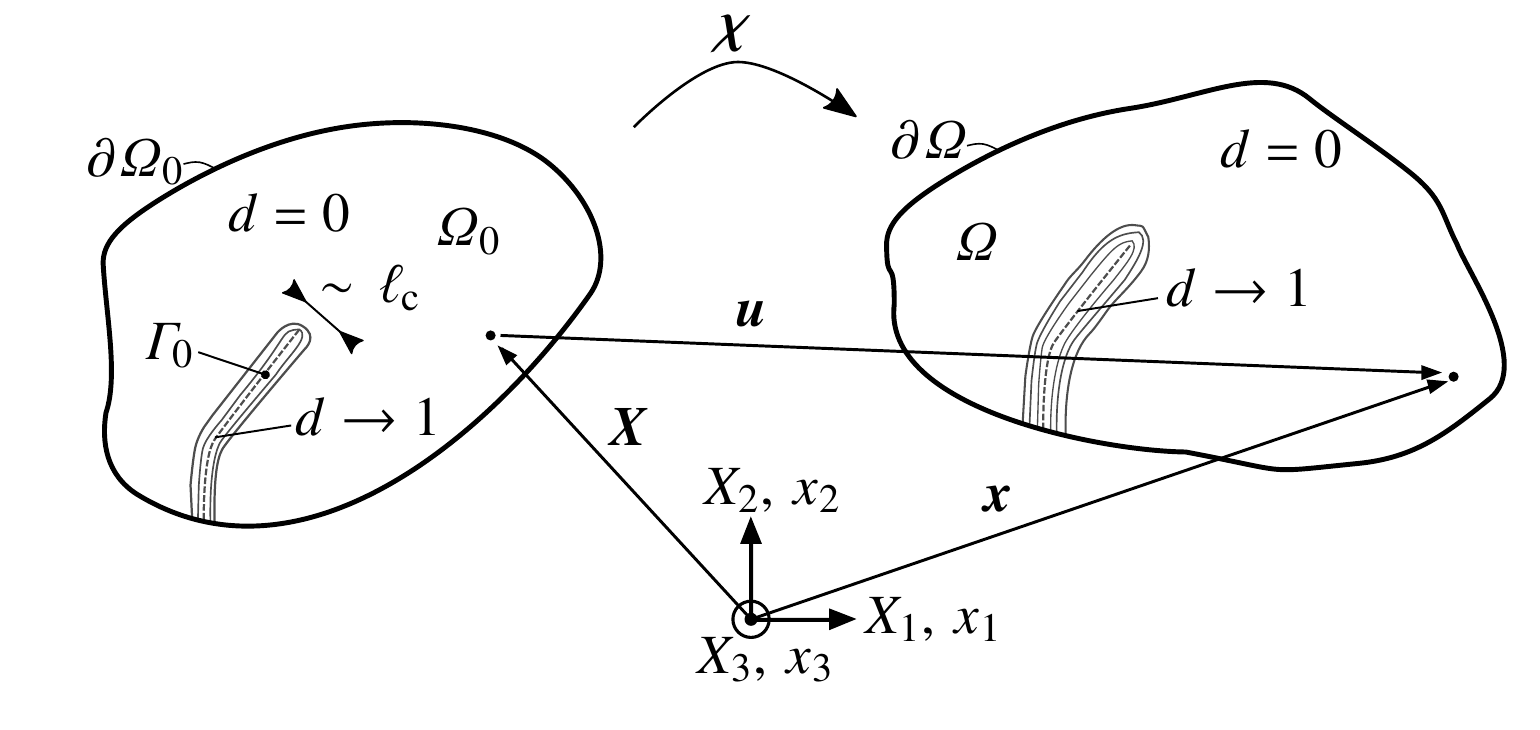} % [trim=left bottom right top]
		\caption{Diffuse representation of a crack within a domain that undergoes finite deformation}
		\label{fig:kinem}
\end{figure}

For a given external load, the deformation of the domain and the crack surface $\varGamma_0$ then can be determined from the equilibrium condition
\begin{equation}
 \varPi \rightarrow \, \nv{stat} \point
\end{equation}
In order to make this energetic approach to fracture accessible to numerical implementation, a regularisation of the functional $\varPi$ is introduced \cite{bourdin2000}.
For this purpose, cracks are described in a diffuse manner by means of a phase-field variable
\begin{equation}
 d: \omref \times [0,t] \rightarrow [0,1] \comma \hspace{3mm} (\ve X, t) \mapsto d(\ve X,t)
 \label{eq:d-def}
\end{equation}
which continuously varies from the intact ($d=0$) to the fully broken ($d=1$) material state.
Using this variable, a \textit{crack surface density}
\begin{equation}
 \gamma_\lc = \frac{1}{4 \, \lc} \left( d^2 + 4 \, \lcs \, \nabr d \cdot \nabr d \right)
 \label{eq:surfDen}
\end{equation} 
can be defined, cf. \cite{miehe2010},%
\footnote{For the crack surface density $\gamma_\lc$, several choices are possible, see e.g. \cite{tanne2018}. The expression adopted here typically is referred to as \mbox{\textit{AT-2}} model\textemdash with reference to the fundamental work of Ambrosio and Tortorelli~\cite{ambrosio1990}.}
in which the regularisation parameter~$\lc$ defines the characteristic width of the diffuse crack and
the Nabla operator with respect to the reference coordinate is defined to
\begin{equation}
 \nabr \, \circ = \sum_{K=1}^N \ve e_K \, \left(\diffp{\, \circ}{X_K}\right) \comma
\end{equation}
with $\ve e_K$ denoting $K$-th basis vector of the Cartesian reference coordinate frame.
Fig.~\ref{fig:kinem} illustrates the concept of diffuse crack representation.
With this approximation of crack surface at hand, dissipation due to crack evolution can be expressed as
\begin{equation}
 \Piflc = \inte{\omref}{\gamma_\lc \, \gc }{V} =: \inte{\omref}{ \Phif }{V} \comma
 \label{eq:fun-risdis-reg}
 \end{equation} 
wherein $\Phif$ is defined as density of fracture pseudo-energy with respect to a volume element in the reference configuration.
In the regularised setting, the decrease of free energy due to fracture is expressed by means of the degradation function
\begin{equation}
 g: [0,1] \rightarrow [0,1] \comma \hspace{3mm} d \mapsto g(d)
 \label{eq:degfu-def}
\end{equation} 
which has to fulfil the conditions
\begin{align*}
 g(d=0)=1 \commam &g(d=1)=0 \comma \\ 
 \frac{\ddp g}{\ddp d} \leq 0 \commam &\left.\frac{\ddp g}{\ddp d}\right|_{d=1}=0 \point
 \numberthis
 \label{eq:degfu-bed}
\end{align*} 
Based upon the degraded reference free energy density
\begin{equation}
 \varPsi = g(d) \, \psi \comma
\end{equation} 
the regularised functional of free energy is given by
\begin{equation}
 \Pistlc = \inte{\omref}{\varPsi}{V} \comma
 \label{eq:free-enfun-reg}
\end{equation}
and the regularised counterpart of the pseudo-energy functional~$\varPi$ reads%
\begin{equation}
 \varPi_\lc = \Pistlc + \Piflc = \inte{\omref}{\varPsi +  \Phif }{V} \point
 \label{eq:fun-reg}
\end{equation}

\paragraph{Generalisation for inelastic material response.}
Following the previous work \cite{dammass2021b} and similar to phase-field fracture models for elasto-plastic materials, 
the free energy density
\begin{equation}
 \varPsi = \gstd \, \psist + \bvi \, \gvid \, \psivi \; =: \Psist + \Psivi
 \label{eq:Psi-inel}
\end{equation} 
is assumed to be additively decomposed into two essential ingredients.%
\footnote{Note that different from \cite{borden2016,kuhn2016b} and in line with e.g. \cite{miehe2010,miehe2016a}, dissipation due to evolution of crack surface is not assumed to contribute to the free energy $\varPsi$, yet included as a distinct contribution $\varPhi$ or $\Piflc$, respectively, to the regularised pseudo-energy functional $\varPi_\lc$.}
Naturally, the first one is the effectively stored strain energy $\Psist$.
In addition, in order to adequately account for the coupling between inelastic deformation and fracture mechanisms, a free energy contribution $\Psivi$ related to accumulated viscous dissipation is assumed.
In the latter, in order to keep the formulation as general as possible, the parameter \mbox{$\bvi \in [0,1]$} is introduced as a weight, cf. \cite{borden2016} and \cite{shen2019}.
The specific definitions of $\psist$ and $\psivi$ considered in this work are given in Sect.~\ref{sec:visco_mod}.%
\footnote{$\psist$ and $\psivi$ can be understood as \textit{virtually undamaged} densities of free energy with respect to a volume element in the reference configuration, i.e. the respective free energy which would be stored in such a reference volume element in the absence of damage.}
For a rigorous motivation and interpretation of $\Psivi$ from a physical point of view, and a numerical investigation in the kinematically linear regime, the reader is referred to~\cite{dammass2021b}.
Contributions similar to $\Psivi$ are also considered in other recent phase-field models of fracture in viscoelastic materials \cite{shen2019,loew2019}.
Furthermore, analogue terms are widely spread in modelling of failure in elasto-plastic materials \cite{borden2016,kuhn2016b,miehe2016a}, where a free energy contribution related to inelastic deformation, which is degraded in case of crack growth can be essential for the description of ductile fracture, cf. \cite{ambati2015,alessi2018a}.%
\footnote{It has to be noted that there are alternative concepts for the phase-field modelling of ductile failure, also. For example, a degradation function~$g$ which, in addition to the fracture phase-field, depends on a measure of plastic deformation \cite{ambati2015,ambati2016b}, and a fracture toughness that diminishes with accumulated inelastic strain \cite{yin2020c}, have been proposed.}

For the two contributions to the free energy, any degradation functions $\gst$ and $\gvi$ satisfying the conditions~\eqref{eq:degfu-bed} can be considered, which, in general, do not have to coincide.
In the literature, different approaches have been taken, e.g. quartic and cubic expressions \cite{kuhn2015,borden2012a}, a sinusoidal ansatz~\cite{yin2020c,yin2020b}, and parametric functions that include additional parameters, which can be fitted to the behaviour of a specific material~\cite{borden2012a,wu2017,sargado2018,steinke2019}.
Without loss of generality, within the scope of this publication, $ \gstd \equiv \gvid \equiv g(d)$ is assumed.
Furthermore, the frequently adopted \cite{bourdin2000,miehe2010,kuhn2016,shen2019} quadratic function
\begin{equation}
 g(d) = (1-k) \, (1-d)^2 + k \comma
 \label{eq:degfu-quadr}
\end{equation} 
in which a small residual $k$ is included in order to enhance numerical stability, is considered. 

\subsection{Viscoelastic bulk response}
\label{sec:visco_mod}

\subsubsection{Kinematics}
\begin{figure}[b!]
		\centering 
		\includegraphics[scale=1.0,trim=0mm 3mm 5mm 0mm]{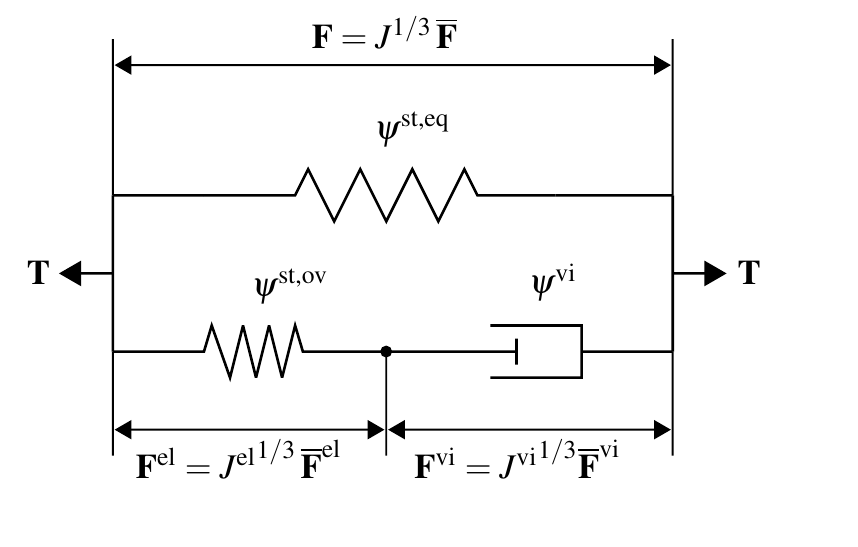} % [trim=left bottom right top]
		\caption{Generalised Maxwell element\textemdash Constitutive assumptions}
		\label{fig:maxwell}
\end{figure}

The displacement of a material point with coordinate \mbox{$\ve X \in \omref$} in the reference configuration is denoted by
\begin{equation}
 \ve u(\ve X,t) = \ve \chi(\ve X,t) - \ve X \comma
\end{equation}
wherein
\begin{equation}
 \ve \chi(\ve X,t) : \omref \times [0,t] \rightarrow \varOmega \comma \hspace{3mm} (\ve X, t) \mapsto \ve x(\ve X,t)
\end{equation} 
is the motion function.
Due to the diffuse approximation of crack topology, $\ve \chi(\ve X,t)$ can be assumed to be bijective and continuous in space and time.
The deformation gradient $\te F$ and its determinant $J$ are then given by
\begin{equation}
 \te F = \left(\nabr \ve \chi\right)^\top \qquad \text{and} \qquad J= \det \, \te F > 0 \point
\end{equation} 
For the rate-dependent deformation behaviour of the material, the approach of Reese and Govindjee~\cite{reese1998} is pursued and a generalised Maxwell model is adopted as shown in Fig.~\ref{fig:maxwell}.%
\footnote{Herein, without loss of generality, only one non-equilibrium branch is considered, which is sufficient for the material investigated in Sect.~\ref{sec:numres}. The extension to multiple non-equilibrium branches can be done in a straightforward manner, though.}
In the non-equilibrium, or over-stress branch, deformation is assumed to consist of an elastic and an inelastic viscous portion, and the deformation gradient is multiplicatively decomposed into
\begin{equation}
 \te F = \fel \cdot \fvi \comma
\end{equation}
accordingly.
Furthermore, following Flory~\cite{flory1961}, a decomposition of the deformation gradient into volumetric and isochoric parts is applied. For the equilibrium branch, the split is given by
\begin{equation}
 \te F = J^{1/3} \, \tei \cdot \fiso \quad ,
 \quad \fiso = J^{-1/3} \, \te F \quad ,
 \quad \det \, \fiso = 1 \comma
 \label{eq:floryTot}
\end{equation}
wherein $\tei$ designates the second-order unit tensor and $\fiso$ is the isochoric portion of the deformation gradient.
 For the the non-equilibrium branch, $\fel$ and $\fvi$ are decomposed separately. Considering, for example, the elastic portion of deformation, its isochoric part is given by
 \begin{equation}
 \feli = {\jel}^{-1/3} \, \fel \comma
 \qquad \det \, \feli = 1 \comma
 \label{eq:floryEl}
 \end{equation}
wherein $\jel = \det \, \fel$.
For the specific definition of the material model, the positive definite left and right Cauchy-Green deformation tensors, $\te b = \te F \cdot \te F^\top$ and $\te C = \te F^\top \cdot \te F$, as well as their elastic counterparts $\bel = \fel \cdot {\fel}^\top$ and $\cel = {\fel}^\top \cdot {\fel}$, are used, respectively.
It has to be noted that $\cel$ does not refer to the current configuration, but to a fictitious \textit{intermediate} configuration defined by $\fvi$. 
The tilde symbol $\tilde \circ$ is introduced to mark quantities which refer to this inelastic intermediate configuration.
Isotropy of the material is assumed and the constitutive equations are specified in terms of principal stretches $\lambda_\alpha$ and $\lame$, which are obtained from the spectral decompositions
\begin{equation}
 \te b = \sum\limits_{\alpha=1}^{N_\lambda} \lambda_\alpha^2 \, \te p_\alpha
 \qquad \text{or} \qquad
 \te C = \sum\limits_{\alpha=1}^{N_\lambda} \lambda_\alpha^2 \, \te P_\alpha
 \comma
\end{equation} 
\begin{equation}
 \bel = \sum\limits_{\beta=1}^{\nlame} \lameq \; \pel
 \qquad \text{or} \qquad
 \cel = \sum\limits_{\beta=1}^{\nlame} \lameq \; \Pel
 \comma
\end{equation} 
in which $N_\lambda \in \{1, 2, 3\}$ and $\nlame \in \{1, 2, 3 \}$ are the number of pair-wise different principal stretches $\lambda_\alpha$ and elastic principal stretches $\lame$, respectively.
The second-order projection tensors, or \textit{eigenvalue-base} tensors, are obtained from
\begin{equation}
 \te p_\alpha = \delta_{1 N_\lambda} \, \te I \, + \prod\limits_{\gamma} \frac{\te b - \lambda^2_\gamma \, \tei}{\lambda^2_\alpha - \lambda^2_\gamma}
\quad , \quad 
\gamma \in [1,N_\lambda]\setminus\alpha \subset \nset
\comma
 \label{eq:projtens}
\end{equation}
with the Kronecker delta $\delta_{\varrho\sigma}$ given by
\begin{equation}
 \delta_{\varrho\sigma} = 
 \left\{ \begin{array}{c}
            1, \quad \varrho = \sigma  \\
            0, \quad \varrho \neq \sigma \\ 
        \end{array}
 \right.
 \comma
 \label{eq:krondelta}
\end{equation}
and equivalent relations for $\te P_\alpha$, $\pel$ and $\Pel$ \cite{miehe1993,miehe1998}.%
\footnote{If there are three pair-wise different principal stretches, i.e. $N_\lambda=3$ or $\nlame=3$, the projection tensors can also be represented by means of the eigenvectors in a straightforward manner, e.g. $\te p_\alpha=\ve n_\alpha \otimes \ve n_\alpha$ with $\ve n_\alpha$ denoting the \mbox{$\alpha$-th} eigenvector of~$\te b$.}

\subsubsection{Specification of free energy densities}

\paragraph{Strain energy.}
The strain energy density is additively decomposed into an equilibrium and over-stress part. Accordingly, for the \textit{virtually undamaged} quantity $\psist$,
\begin{equation}
 \psist = \psiste(\te C) + \psisto(\te C, \fvi )
\end{equation} 
is defined, wherein $\te C$ and $\fvi$ form the set of independent thermodynamic state variables considered here, in addition to the phase-field variable $d$.
Each contribution splits further into a volumetric portion $\psistev$ and $\psistov$, and an isochoric part $\psistei$ and $\psistoi$, respectively.
A compressible Ogden model \cite{ogden1997} is assumed for both the equilibrium and non-equilibrium branches and the respective strain energy density contributions are defined to
\begin{align*}
 \psiste =& \psistev +\psistei \\
 =& \frac{\keq}{4} \left(J^2 - 2 \ln J -1 \right) \\
 +& \sum_{p=1}^{\Neq} \frac{\mueq}{\aleq} \, 
		\left(
		\sum_{\varrho=1}^{N_\lambda} \nu_\varrho \, \bar \lambda_\varrho^{\aleq} -3
        \right)
        \comma
        \numberthis
        \label{eq:psiste}
\end{align*}
% ov-contribution
\begin{align*}
 \psisto =& \psistov +\psistoi \\
 =& \frac{\kov}{4} \left({\jel}^2 - 2 \ln \jel -1 \right) \\
 +& \sum_{p=1}^{\Nov} \frac{\muov}{\alov} \, 
		\left(
		\sum_{\sigma=1}^{\nlame} \nulame[\sigma] \, \left(\lamei \right)^{\alov} -3
        \right)
        \comma
        \numberthis
        \label{eq:psisto}
\end{align*}
wherein $\bar \lambda_\varrho = J^{-1/3} \, \lambda_\varrho$ and $\lamei[\sigma] = \jel^{-1/3} \, \lame[\sigma]$ are the isochoric total and elastic principal stretches following from~ \eqref{eq:floryTot} and \eqref{eq:floryEl}. Their algebraic multiplicity is given by $\nu_\varrho \in \{1, 2, 3\}$ and $\nulame[\sigma] \in \{1, 2, 3\}$, respectively.
Furthermore, the compression moduli are denoted by $\keq >0$ and $\kov >0$, and $\Neq$, $\aleq$, $\mueq > 0$, as well as $\Nov$, $\alov$, $\muov >0$ are parameters of the Ogden models.
% mu, nu
From these constants, the initial shear moduli and the according Poisson's ratios can be defined. For example, for the equilibrium branch, they read
\begin{equation}
 \mueqn = \frac{1}{2} \sum\limits_{p=1}^{\Neq} \mueq \, \aleq 
 \quad \text{and} \quad
 \nueqn = \frac{3 \, \keq - 2 \, \mueq}{2 \, (3 \, \keq + \mueq)}
 \comma
\end{equation} 
and similar relations hold for the non-equilibrium branch.

\paragraph{Viscous contribution.}
The degraded free energy contribution $\Psivi$ related to inelastic mechanisms is designed such that a certain portion of accumulated viscous dissipation can enter the phase-field fracture driving force.
Before defining the respective \textit{virtually undamaged} quantity $\psivi$ in the finite viscoelasticity framework, the simple setting of a uniaxial deformation in the kinematically linear regime is considered for motivational purpose.
Then, in the absence of damage, viscous dissipation in a material described by means of the generalised Maxwell model takes the form
\begin{equation*}
 \disua = \binte{0}{t}{\eta \, {\dot{\varepsilon}^\nv{vi}} \, \, {\dot{\varepsilon}^\nv{vi}}}{\bar t}
 \comma
\end{equation*}
in which ${\dot{\varepsilon}^\nv{vi}}$ is the rate of inelastic deformation and $\eta$ designates the viscosity of the material.
In order to generalise $\disua$, the tensor
\begin{equation}
 \dvie = - \frac{1}{2} \,\lie\left[ \bel \right] \cdot {\bel}^{-1} \comma
 \label{eq:dvie-def}
\end{equation} 
is introduced as a measure of the rate of inelastic deformation in the finite viscoelasticity setting, wherein
\begin{equation}
 \lie\left[ \bel \right] = \te F \cdot \dot{\left( \cvi^{-1} \right)} \cdot \te F^\top
 \quad \text{with} \quad 
 \cvi = \fvi^\top \cdot \fvi
 \label{eq:lie-def}
\end{equation} 
is the Lie derivative of $\bel$.
Furthermore, a fully symmetric, positive definite, isotropic fourth-order tensor 
\begin{equation}
 \visc = 2 \, \etai \, \tte I^\text{D} + 9 \, \etav \, \te I \otimes \te I
 \label{eq:visc-def}
\end{equation} 
is defined, in which $\tte I^\text{D}$,
\begin{equation}
\tte I^\text{D}_{klmn} = \frac{1}{2} \left( \delta_{km} \delta_{ln}+\delta_{kn} \delta_{lm} \right) - \frac{1}{3}\delta_{kl} \delta_{mn}
\comma
\end{equation}
is the fully symmetric fourth-order deviator projection tensor. Therein, $\etai, \, \etav > 0$ are viscosities with respect to the isochoric and volumetric portion of deformation, respectively.
The \textit{virtually undamaged} free energy density contribution related to viscous mechanisms is then defined to
\begin{equation}
 \psivi = \binte{0}{t}{\dvie \, \colon \visc \, \colon \dvie}{\bar t} \comma
\end{equation} 
which is positive and monotonically increasing in time.

\paragraph{Remark on the measure of rate of inelastic deformation.} 
The definition of $\dvie$ \eqref{eq:dvie-def} can be written in an alternative form, which may be more intuitive.
For this purpose, the inelastic velocity gradient
\begin{equation}
 \lvit = \fvid \cdot \fvi^{-1}
 \label{eq:lvit-def}
\end{equation}
is introduced. It refers to the intermediate configuration defined by $\fvi$.
The counterpart of $\lvit$ transformed to the current configuration reads
\begin{equation}
 \lvi = \fel \cdot \lvit \cdot \fel^{-1} \point
 \label{eq:lvi-def}
\end{equation}
Assuming that there is no inelastic spin, i.e. $\lvit =  \dvit $ with $\dvit = \sym \lvit$ denoting the rate of inelastic deformation with respect to the viscous intermediate configuration,
\eqref{eq:lie-def}\textsubscript{1} can be rewritten as
\begin{equation}
 \lie\left[ \bel \right] = - 2 \, \fel \cdot \dvit \cdot \fel^\top 
 \label{eq:lie-def-altern}
\end{equation} 
and
\begin{equation}
  \fel \cdot \dvit \cdot \fel^{-1}
  =
  - \frac{1}{2} \,\lie\left[ \bel \right] \cdot {\bel}^{-1}
  =\dvie
\end{equation} 
holds, from which, together with the transformation rule \eqref{eq:lvi-def}, the definition of $\dvie$ as an Eulerian measure of rate of inelastic deformation becomes clear.
For more details, the reader is referred to \cite{wriggers2008}, where similar kinematic relations are derived in the context of plasticity.

\subsection{Evolution of phase-field}
\label{sec:pf-evol}

The equation governing the evolution of the fracture phase-field variable is deduced from the pseudo-energy functional $\Pilc$ by means of the variational derivative
\begin{equation}
 \diffv{\varPi_\lc}{d} = - \etaf \, \dot{d}
 \hspace{10mm} \nv{and} \hspace{10mm} 
 \left. \nabr d \cdot \ve N \right|_{\partial \, \omref} = 0 \comma
 \label{eq:pfgl-allg}
\end{equation}
wherein $\etaf$ is introduced as a kinetic fracture parameter in order to avoid discontinuity of the field variables in time and for numerical purposes, i.e. for enhancing the stability of the solution scheme, cf. \cite{gurtin1996,kuhn2016} and $\ve N$ denotes the outward-pointing unit normal vector on $\partial \omref$.
For the simulations presented in Sect.~\ref{sec:numres}, $\etaf$ is chosen such small that its influence on the simulation results vanishes which is verified by means of a comparative study of different values.%
\footnote{In several other models, e.g. \cite{miehe2014,loew2019}, $\etaf$ is assigned a finite value and thus considered as an additional material parameter.
On the one hand, such a direct coupling of rate effects into the evolution of phase-field by means of $\dot d$ can enable more modelling flexibility especially regarding the post-critical stage of a response.
On the other hand, when it comes to damage, incorporation of a finite $\etaf$ is equivalent to assuming a pseudo-viscous dissipation in addition to proper viscous effects and fracture dissipation. 
However, for fracture dissipation, according to the fundamental modelling hypothesis~\eqref{eq:fun-nireg}, the fracture toughness $\gc$ is assumed to be the essential parameter.
Therefore, a toughness depending on rate of deformation is presumed to be more consistent from an energetic point of view if a direct coupling of rate-effects into phase-field evolution is necessary.
Furthermore, a finite $\etaf$ would also incorporate some redundant information which should rather be taken into account by the viscoelastic model for deformation.}

Inserting the definitions made in the previous Sections into $\varPi_\lc$, the evolution equation \eqref{eq:pfgl-allg}\textsubscript{1} takes the form
\begin{align*}
  - \etaf \, \dot{d} =& \diffp{g}{d} \left( \psist + \bvi \, \psivi \right) \\
  &+ \gc \left( \frac{1}{2 \, \lc} d - 2\, \lc \, \nabr \cdot \nabr d \right)
  \numberthis \label{eq:pfgl-unmod} 
\end{align*}
from which it becomes clear that, depending on the specific choice of $\bvi$, fracture is driven by stored strain energy and the free energy contribution related to a portion of accumulated viscous dissipation.
It has to be noted that in the present form~\eqref{eq:pfgl-unmod}, the evolution equation enables the phase-field variable to decrease, i.e. crack healing is not prohibited.
Therefore, a modification is adopted which overcomes this issue, see Sect.~\ref{sec:irr}.

\subsection{Rate-dependent fracture toughness}
\label{sec:gc_def}

\begin{figure}[tb]
		\centering 
		\includegraphics[trim=10mm 0 0 0,width=0.33\textwidth]{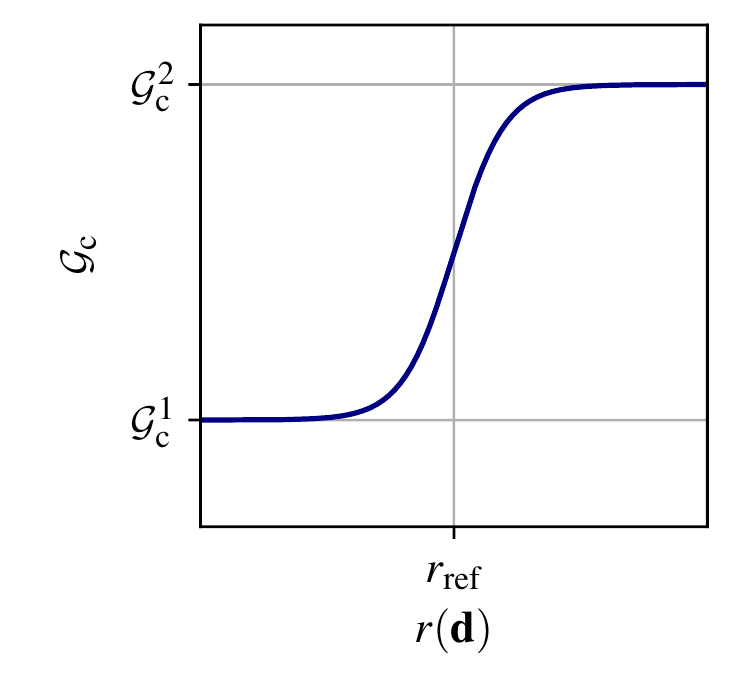} % [trim=left bottom right top]
		\caption{Rate-dependent fracture toughness function $\gc\left(r(\te d)\right)$}
		\label{fig:gc_d}
\end{figure}

For various materials, in addition to or instead of the deformation behaviour of the bulk material, the resistance against fracture has been reported to depend on rate of deformation.
For instance, in elastomers, at low rates of deformation, chain entanglements can be resolved, which is not the case at high rates of deformation. Therefore, the number of chemical bonds that are broken when a crack propagates can be assumed to rise with rate of deformation and the fracture toughness increases accordingly, cf. \cite{gent1994,gent2012} for a more detailed discussion and experimental results.
Furthermore, for several natural materials and foods, where the underlaying microscopic mechanisms can be more complex, a rate-dependency of $\gc$ has been reported \cite{goh2005,gamonpilas2009,forte2015}.%
\footnote{It has to be noted that, unlike here, in some publications dealing with fracture of inelastic materials, $\gc$ is not only regarded as a measure of dissipation directly coming along with breaking of bonds but also comprises dissipative mechanisms of the bulk material.}
Therefore, as an extension of the variational phase-field equation \eqref{eq:pfgl-unmod}, fracture toughness is considered to depend on deformation rate, which enables a maximum of flexibility in modelling rate-dependent fracture processes.
For this purpose, 
\begin{equation}
 r (\te d) = \left\Arrowvert \te d \right\Arrowvert_\nv{F} = \sqrt{\te d \, \colon \te d}
\end{equation}
is introduced as a scalar measure of effective rate of deformation.
Furthermore, without loss of generality, in line with \cite{miehe2015b}, the sigmoid-shaped function
\begin{equation}
 \gc(\te d) = \frac{\gco + \gct}{2} + \left( \frac{\gct - \gco}{2} \right)
 \, \tanh \left[ c \cdot (r(\te d) - \rref )\right]
 \label{eq:rabGc-def}
\end{equation}
is adopted, see Fig.~\ref{fig:gc_d}.
The extended phase-field evolution equation can then be written as
\begin{align*}
  - \etaf \, \dot{d} =& \diffp{g}{d} \left( \psist + \bvi \, \psivi \right) \\
  &+ \gc(\te d) \left( \frac{1}{2 \, \lc} d - 2\, \lc \, \nabr \cdot \nabr d \right)
  \point
  \numberthis \label{eq:pfgl-unmod-gcd} 
\end{align*}
In Sect.~\ref{sec:rabGc}, for different parameters $\gco$, $\gct$, $c$, $\rref$, coupling between rate-dependent deformation and toughness is analysed.
\paragraph{Fracture pseudo-energy and rate-dependent toughness.}
If fracture toughness is a function of rate of deformation and thus implicitly depends on time, density of fracture pseudo-energy $\Phif$ has to be rewritten as
\begin{equation}
 \Phifd = \dot \gamma_\lc \, \gc(\te d)  \quad \Leftarrow \quad \Phif = \binte{0}{t}{\dot \gamma_\lc \, \gc\left(\te d[\bar t]\right))}{\bar t}
\end{equation} 
in order to account for possible a posteriori changes of $\gc$ following the evolution of fracture phase-field at a given time~$\bar t$.
Therefore, the phase-field equation~\eqref{eq:pfgl-unmod-gcd} can be seen as a non-variational extension of \eqref{eq:pfgl-unmod}, similar to the suggestions made in \cite{schaenzel2015,bilgen2019b}, for instance.
Furthermore, it is noted that, different from e.g. \cite{yin2020g}, in the proposed model, rate-dependency of fracture toughness does not affect the density of free energy $\varPsi$, since dissipation due to crack growth is not supposed to enter $\varPsi$. Accordingly, no additional stress terms arise from rate-dependent toughness, see the evaluation of the entropy inequality below in Sect.~\ref{sec:thermocons}.

\subsection{Stress tensor, viscous evolution and thermodynamic consistency}
\label{sec:thermocons}

Under isothermal conditions, the second law of thermodynamics can be stated by means of the density of dissipation power $\disd$ as
\begin{equation}
 \disd = \frac{1}{2} \, \te T \, \colon \dot{\te C} - \dot \varPsi \geq 0 \comma
\end{equation} 
cf.~\cite{coleman1963}, with $\te T$ denoting the second Piola-Kirchhoff stress tensor.
For $\varPsi = \varPsi(\te C, \fvi, d)$, this inequality can be rewritten to
\begin{align*}
\disd & =  \left( \frac{1}{2} \, \te T - g(d) \, \diffp{\psist}{\te C} \right) \, \colon \dot{\te C} 
   \underbrace{- \diffp{g}{d} \left( \psist + \bvi \psivi \, \right) \dot d}_{\dpfd} \\
  & \underbrace{- g(d) \left( \diffp{\psist}{\fvi} \, \colon \dot{\fvi} + 
  \bvi \, \dvie \, \colon \visc \, \colon \dvie \right)}_{\dvid}
  \geq 0
  \numberthis
  \comma
\end{align*}
wherein the contributions to dissipation power density due to fracture, $\dpfd$, and viscous effects, $\dvid$, can be identified.
The standard argument that $\disd \geq 0$ shall hold for arbitrary processes leads to the definition of stress
\begin{equation}
 \te T = 2 \, g(d) \, \diffp{\psist}{\te C}
 =
  2 \, g(d) \, \left( 
 \underbrace{\diffp{\psiste}{\te C}}_{\ute/2}
 +
 \underbrace{\diffp{\psisto}{\te C}}_{\uto/2} 
 \right)
 \comma
 \label{eq:PK2stressA}
\end{equation} 
with the \textit{virtually undamaged} equilibrium and over-stress tensors denoted by $\ute$ and $\uto$, respectively, and the residual inequalities
\begin{equation}
 \dpfd \geq 0 \glmand \dvid \geq 0 \point
\end{equation} 
\paragraph{Stress tensor.}
Inserting the definitions of $\psiste$ and $\psisto$, Eqs. \eqref{eq:psiste} and \eqref{eq:psisto}, into \eqref{eq:PK2stressA}, the contributions to the second Piola-Kirchhoff stress tensor take the form
\begin{equation}
		\begin{split}
			\ute = \sum_{\beta=1}^{N_\lambda} \,
			&\frac{1}{\lambda_\beta^2} \,
			\Biggl[ \,
			\sum_{p=1}^{\Neq} \mueq \, \left(
			\bar \lambda_\beta^{\aleq} -\frac{1}{3} \sum_{\varrho=1}^{N_\lambda}
			\nu_\varrho \, \bar\lambda_\varrho^{\aleq}\right) 
			\\
			&+ \frac{\keq}{2} \, (J^2-1)
			\Biggr] \, \te P_\beta
			\comma
		\end{split}
 \label{eq:PK2stressEq}
\end{equation}
\begin{equation}
		\begin{split}
			\uto = 
			\fvi^{-\top} \cdot
			\Biggl\{ \,
			\sum_{\beta=1}^{\nlame} &\, \frac{1}{\left[\lame[\beta]\right]^{2}} \,
			\Biggl[
			 \, \sum_{p=1}^{\Nov} \muov \, 
			\Biggl(
                \left[ \lamed \right]^{\alov} 
                \\
                &-\frac{1}{3} \sum_{\varrho=1}^{\nlame}
                \nulame[\sigma] \, \left[\lamed[\varrho]\right]^{\aleq}
			\Biggr) 
			\\
			&+ \frac{\kov}{2} \, ({\jel}^2-1)
			\Biggr] \, \Pel
			\Biggr\}
			\cdot \fvi^{-1}
			\point
		\end{split}
 \label{eq:PK2stressOv}
\end{equation} 
\paragraph{Residual inequalities.}
As both $\psist$ and $\psivi$ are positive, {$\bvi \in [0,1]$}, and due to \eqref{eq:degfu-bed}$_3$, the condition $\dpfd \geq 0$ reduces to $\dot d > 0$, i.e. irreversibility of fracture. The fulfilment of this demand will be addressed in Sect.~\ref{sec:irr}.

Due to \eqref{eq:degfu-def}, $\dvid \geq 0$ reduces to
\begin{equation}
 \diffp{\psist}{\fvi} \, \colon \dot{\fvi} + 
 \bvi \, \dvie \, \colon \visc \, \colon \dvie \leq 0 \point
 \label{eq:dvid2}
\end{equation}
Making use of the relations outlined in Sect.~\ref{sec:visco_mod}, after some lengthy manipulations, the first term can be rewritten as
\begin{equation}
 \diffp{\psist}{\fvi} \, \colon \dot{\fvi} = \utao \, \colon \dvie \comma
\end{equation} 
wherein
\begin{equation}
 \utao = 2 \, \diffp{\psist}{\bel} \cdot \bel = \te F \cdot \uto \cdot \te F^\top
\end{equation} 
is the \textit{virtually undamaged} Kirchhoff over-stress. For a more detailed derivation see also \cite{reese1998}.
Then \eqref{eq:dvid2} takes the form
\begin{equation}
 \left(-\utao + \bvi \, \visc \, \colon \dvie \right) \, \colon \dvie \leq 0 \comma
 \label{eq:dvi-ungl-mod}
\end{equation} 
from which, in line with \cite{reese1998}, the equation governing viscous evolution
\begin{equation}
 \utao = \visc \, \colon \dvie
 \label{eq:def-visc-ev}
\end{equation} 
is defined. By reason of $\bvi \in [0,1]$, the quadratic form obtained from inserting \eqref{eq:def-visc-ev} into \eqref{eq:dvi-ungl-mod} is compatible with the second law of thermodynamics.

\section{Algorithmic aspects}
\label{sec:impl}

\subsection{Irreversibility of fracture}
\label{sec:irr}

In order to guarantee irreversibility of fracture, the history variable approach of Miehe et al.~\cite{miehe2010} is pursued.
For this purpose, the phase-field equation~\eqref{eq:pfgl-unmod-gcd} is rewritten to
\begin{align*}
  - \frac{\etaf}{\gc} \, \dot{d} =& \diffp{g}{d} \, \mathcal H 
  +  \left( \frac{1}{2 \, \lc} d - 2\, \lc \, \nabr \cdot \nabr d \right)
  \comma
  \numberthis \label{eq:pfgl-mod} \\ 
\end{align*}
wherein the history variable 
\begin{equation}
 \hist = \max_{\tau \in [0,t]} \left\{\frac{1}{\gc(\te d(\tau))} \, \psist(\tau) + \bvi \, \psivi(\tau) \right\}
 \label{eq:histvar}
\end{equation} 
comprises the maximum of \textit{virtually undamaged} fracture driving force which has occurred.
With this form of the phase-field evolution at hand, the governing equations of the model are summarised in Tab.~\ref{tab:gov-eqs} considering the total Lagrangian approach.
\begin{table}[tb]
\caption{Governing equations for the present model following the total Lagrangian approach: Balance of linear momentum~(a), phase-field equation~(b), viscous evolution~(c), rate-dependent fracture toughness~(d). Without loss of generality, volume forces are neglected in~(a).}
\centering
\resizebox{\linewidth}{!}{
 \begin{tabular}{p{\linewidth}}
 \hline
 \[ \nabr \cdot (\te T \cdot \te F^\top)= \ve 0 \tag{a} \]\\
{\begin{align*} 
  - \frac{\etaf}{\gc} \, \dot{d} =& \diffp{g}{d} \, \mathcal H 
  +  \left( \frac{1}{2 \, \lc} d - 2\, \lc \, \nabr \cdot \nabr d \right) \tag{b} \\
\quad \text{with} \quad \hist =& \max_{\tau \in [0,t]} \left\{\frac{1}{\gc(\te d(\tau))} \, \psist(\tau) + \bvi \, \psivi(\tau) \right\}
\end{align*}} \\
 \[ \utao = \visc \, \colon \dvie \quad \text{with} \quad  \dvie = - \frac{1}{2} \,\lie\left[ \bel \right] \cdot {\bel}^{-1} \tag{c} \]\\
{\begin{align*} 
\gc(\te d) &= \frac{\gco + \gct}{2} + \left( \frac{\gct - \gco}{2} \right) \, \tanh \left[ c \cdot (r(\te d) - r_\text{ref} ) \right] \tag{d}\\
& \quad \text{with} \quad r (\te d) = \left\Arrowvert \te d \right\Arrowvert_\nv{F}
\end{align*}} \\
  \hline
  \end{tabular}}
\label{tab:gov-eqs} 
\end{table}
Alternatively, in line with \cite{kuhn2010,kuhn2013}, Dirichlet boundary conditions can be applied to the phase-field on all nodes
\begin{equation}
 \xirr \in \left\lbrace \ve X \in \omref \; \left| \; \exists \, \tau \in [0,t] : d(\ve X,\tau) \geq d_\nv{crit} \right. \right\rbrace
\end{equation} 
where the phase-field variable has reached a critical value $d_\nv{crit}$:
\begin{equation}
 d(\xirr) \muss 1 \; \forall \; \xirr \point
 \label{eq:fr-irr}
\end{equation} 
For the setups analysed in Sect.~\ref{sec:numres}, the two approaches have been compared, exemplary, and no relevant differences could be noticed.%
\footnote{For the simulation of relaxation-dominated load cases together with $\bvi > 0$, special attention has to be paid to the fact that $\psivi$ is incorporated into $\mathcal H$ according to \eqref{eq:histvar}.
Therefore, in these cases, either an altered definition of the history variable or the Dirichlet boundary condition approach would be more reasonable, cf. \cite{dammass2021b}.}

\subsection{Viscous evolution}
\label{sec:vis-evol}

For the integration of viscous evolution equation \eqref{eq:def-visc-ev}, an operator split scheme of \textit{predictor-corrector} type is adopted as proposed in \cite{reese1998}.
Within the scope of this well-established approach, the evolution of elastic deformation
\begin{equation}
 \dot{\te b}^\nv{el} = \underbrace{\te l \cdot \bel + \bel \cdot \te l^\top}_\nv{predictor} +  \underbrace{\te F \cdot \dot{\left( \cvi^{-1} \right)} \cdot \te F^\top}_\nv{corrector}
 \comma
 \label{eq:bel-evol}
\end{equation} 
is split into the contributions from change in total deformation, which is considered in the predictor step, and viscous evolution, which is accounted for in the inelastic corrector. For the predictor step, viscous deformation $\fvi$ or $\cvi$ is frozen, giving a \textit{trial} state of elastic deformation at time step~$t_n$ to
\begin{equation}
 {}_n\beltr = {}_n\te F \cdot {}_{n-1}\cvi^{-1} \cdot {}_n\te F^\top \point
\end{equation}
Subsequently, within the corrector step, \eqref{eq:bel-evol} is evaluated for the total deformation assumed to be constant, i.e. $\te l= \te 0$, which, with evolution equation \eqref{eq:def-visc-ev} and kinematic relations \eqref{eq:dvie-def} and \eqref{eq:lie-def} can then be written as
\begin{equation}
 \dot{\te b}^\nv{el} = -2 \, \visc^{-1} \, \colon \left( \utao \cdot \bel \right)
 \point
 \label{eq:bel-evol-corr}
\end{equation}
Due to isotropy, the principal directions of $\bel$, $\beltr$ and $\utao$ coincide, which makes the evaluation of \eqref{eq:bel-evol-corr} in terms of elastic principal stretches $\lame$ attractive. 
For the viscosity tensor $\visc$ defined according to \eqref{eq:visc-def}, this leads to
\begin{equation}
 \frac{\nv{d}}{\nv{d}\,t}\left(\lame\right)^2 = - \left[\frac{1}{\etai} \, \utaodev + \frac{2}{9 \, \etav} \tr \utao  \right] \, \left(\lame\right)^2 \comma
 \label{eq:visc_corr_interm}
\end{equation} 
wherein $\utaodev$ denote the principal components of the over stress deviator $\dev \utao$.
Within the scope of the FE framework, differential equation \eqref{eq:visc_corr_interm} is integrated in an approximate manner by means of an exponential mapping ansatz and rewritten in terms of logarithmic elastic principal stretches $\epse = \ln \lame$ as
\begin{equation}
 0 = \epse + \Delta t \left[\frac{1}{2 \, \etai} \, \utaodev + \frac{1}{9 \, \etav} \tr \utao \right] - \epsetr =: r_\beta
 \point
 \label{eq:lokit-gls-3d}
\end{equation}
Generally, $\epse$ are determined from an iterative solution of the system of non-linear algebraic equations $r_\beta = 0$ with \mbox{$\beta \in [1,N] \subset \mathbb N$}.
However, in case of two-dimensional plane stress setups as considered in Sect.~\ref{sec:numres}, in addition to these three equations, it has to be ensured that the out of plane stresses vanishes, i.e. $\utaoj[3]=0$ must hold.
In these cases, in addition to $\epse$, the out of plane stretch $\lambda_3$ has to be determined from the system of equations
\begin{equation}
 w_\varrho := \left[\utaoj[3] \quad r_1 \quad r_2 \quad r_3 \right]^\top = 0 \point
 \label{eq:lokit-gls-esz}
\end{equation}
Regardless of whether a plane stress state is considered or not, the respective system of equations \eqref{eq:lokit-gls-3d} or \eqref{eq:lokit-gls-esz} is solved by means of a \textit{local} Newton iteration scheme at each quadrature point.
In the following, the procedure is briefly described for the case that a plane stress state has to be guaranteed.
With the vector of unknowns then written as
\begin{equation}
  x_\sigma := \left[\lambda_3 \quad \epse[1] \quad \epse[2] \quad \epse[3] \right]^\top
\end{equation} 
and the \textit{local} tangent matrix
\begin{equation}
 K_{\varrho \sigma} = \diffp{w_\varrho}{x_\sigma} \comma
\end{equation}
the linearisation of \eqref{eq:lokit-gls-esz} around ${}^j x_\sigma$ is given as
\begin{equation}
 \prescript{j}{n}w_\varrho \approx \sum_{\sigma=1}^4 \prescript{j-1}{n}w_\varrho + \left.K_{\varrho \sigma}\right|_{\prescript{j-1}{n}x_\sigma} \, \left(\prescript{j}{n}x_\sigma - \prescript{j-1}{n}x_\sigma \right) = 0
 \point
\end{equation} 
For the specification of the derivatives $\partial{w_\varrho}/\partial{x_\sigma}$, the reader is referred to Appendix \ref{sec:app_loctan}.
Based on the linearisation, the Newton procedure is carried out as summarised in Algorithm box~\ref{alg:locit-esz}.
For this, at each increment $t_n$, the iteration scheme is initialised by means of
\begin{equation}
 \prescript{j=0}{n}x_\sigma = \left[{}_{n-1}\lambda_3 \quad {}_{n}\epsetr[1] \quad {}_{n}\epsetr[2] \quad \left.\epsetr[3]\right|_{{}_{n-1}\lambda_3} \right]^\top
\end{equation} 
with
\begin{equation}
 \left.\epsetr[3]\right|_{{}_{n-1}\lambda_3} = \prescript{}{n-1}\varepsilon_3 + \frac{1}{2} \, \ln\left[{}_{n-1}\cvi^{-1}_{33} \right] \point
\end{equation} 
Within the iterative solution procedure, special attention has to be paid to $_\nv{tr}\epse[3]$ as it needs to be updated after each \textit{local} iteration $j$ according to 
\begin{equation}
\prescript{j}{n} \epsetrdr= \prescript{j}{n}\varepsilon_3 + \frac{1}{2} \, \ln\left[{}_{n-1}\cvi^{-1}_{33} \right]
\end{equation} 
due to the change of $\varepsilon_3 = \ln \lambda_3$.
\begin{algorithm}[tb]
\hrulefill

\small
 Initialisation at each increment $n$:
\[ j=0 \comma \]
\[ \prescript{j=0}{n}x_\sigma = \left[{}_{n-1}\lambda_3 \quad {}_{n}\epsetr[1] \quad {}_{n}\epsetr[2] \quad \left.\epsetr[3]\right|_{{}_{n-1}\lambda_3} \right]^\top \comma\]
\[ \prescript{j=0}{n}w_\varrho = \left.w_\varrho\right|_{\prescript{j=0}{n}x_\sigma}
\quad , \quad
\prescript{j=0}{n}K_{\varrho \sigma}=\left.K_{\varrho \sigma}\right|_{\prescript{j=0}{n}x_\sigma} 
\]
 \While{$\|\prescript{j}{n}w_\varrho\|_\infty > \nv{tol}$}{
    Solution of the linearised system of equations:
        \[\prescript{j+1}{n}x_\sigma = - \sum_{\sigma=1}^4 \prescript{j}{n}K^{-1}_{\sigma \varrho} \, \prescript{j}{n}{w_\varrho} + \prescript{j}{n}x_\sigma \]
    Update of dependent quantities:
    \[\prescript{j+1}{n}{\epsetr[3]} \quad , \quad \prescript{j+1}{n}w_\varrho \quad , \quad 
    \prescript{j+1}{n}{K} %= \left.{K}_{\varrho \sigma}\right|_{\prescript{j+1}{n}x_\sigma}
    \]
    \[ j:=j+1\]
 }
 \hrulefill
 \vspace{3mm}
 \caption{\textit{Local} Newton iteration scheme in case of plane stress state}
 \label{alg:locit-esz}
\end{algorithm}

\subsection{Weak forms of the governing equations}
\label{sec:fe-impl}

For the derivation of the weak forms of equilibrium and phase-field equation, the test function spaces
\begin{align*}
 \wusj & := \left\lbrace \wuj \in \mathbb{H}^1(\omref) \,  \left| \, \wuj = 0 \; 
  \forall \, \ve X \in \domrefuj \right. \right\rbrace \comma \\
  & \hspace{2cm} j \in [1,N] \subset \mathbb N \comma
 \numberthis
\end{align*} 
and
\begin{equation}
 \wcs = \mathbb{H}^1(\omref)
\end{equation} 
are defined.
Therein, $\mathbb{H}^1(\omref)$ is the Sobolev space of square integrable functions possessing square integrable derivatives in $\omref$, and $\domrefuj$ denotes the parts of the boundary where the $j$-component of the displacement vector $\ve u$ is prescribed.
Then, (a) and (b) from Tab.~\ref{tab:gov-eqs} are multiplied by
\begin{equation}
 \wu = \left[ \wuj[1] \, \cdots \, \wuj[N] \right]^\top, \quad \wuj \in \wusj
\comma
\end{equation} 
and $\wc \in \wcs$, respectively. Integration by parts and making use of the divergence theorem yields
\begin{equation}
  \inte{\omref}{\left(\te T \cdot \te F^\top\right) \, \colon \left(\nabr \wu\right)^\top}{V} 
  - \inte{\partial \omref}{\hat{\ve p} \, \wu}{A} = 0 \comma
  \label{eq:impb-s}
\end{equation} 
wherein $\hat{\ve p}$ denotes the Piola traction vector with its components $\hat p_j$ prescribed on $\partial \omref \setminus \domrefuj$, and
\begin{equation}
\begin{aligned}
 & \inte{\omref}{ \left( \diffp{g}{d} \, \mathcal H  + \frac{1}{2 \, \lc} \, d + \frac{\etaf}{\gc} 
 \, \dot d \right) \, \wc \\
 & + 2 \, \lc \, \nabr d \cdot \nabr \wc }{V}  
 +  \inte{\partial \omref}{\, 2 \, \lc \, \underbrace{\nabr d \cdot \ve N}_{= \, 0,\text{ cf. \eqref{eq:pfgl-allg}}} \, \wc }{A} 
 = 0
 \point
 \end{aligned}
 \label{eq:pfgl-s}
\end{equation}
Time discrete forms are obtained by approximating the respective rates using an Euler backward scheme.
For spatial discretization, Galerkin's method is applied. Then, the discretized equations are implemented into a standard finite element framework.
The coupled problem is solved by means of a \textit{staggered} approach. Furthermore, adaptive control of the time step size is employed based on a heuristic scheme.
Information on the material tangent that is required for the iterative solution of \eqref{eq:impb-s} is given in Appendix~\ref{sec:app_mattan}. 

\section{Representative simulations}
\label{sec:numres}

In this Section, several numerical examples are presented
in order to analyse the characteristics of the present model and to demonstrate its flexibility in describing different responses.
Furthermore, the comparison of numerical predictions to experimental results of Loew et al.~\cite{loew2019} serves for validation of its predictive capabilities.

\subsection{Parameter identification from experimental data}
\label{sec:parident}

\begin{figure}[tb]
		\centering 
		\includegraphics[width=0.45\textwidth]{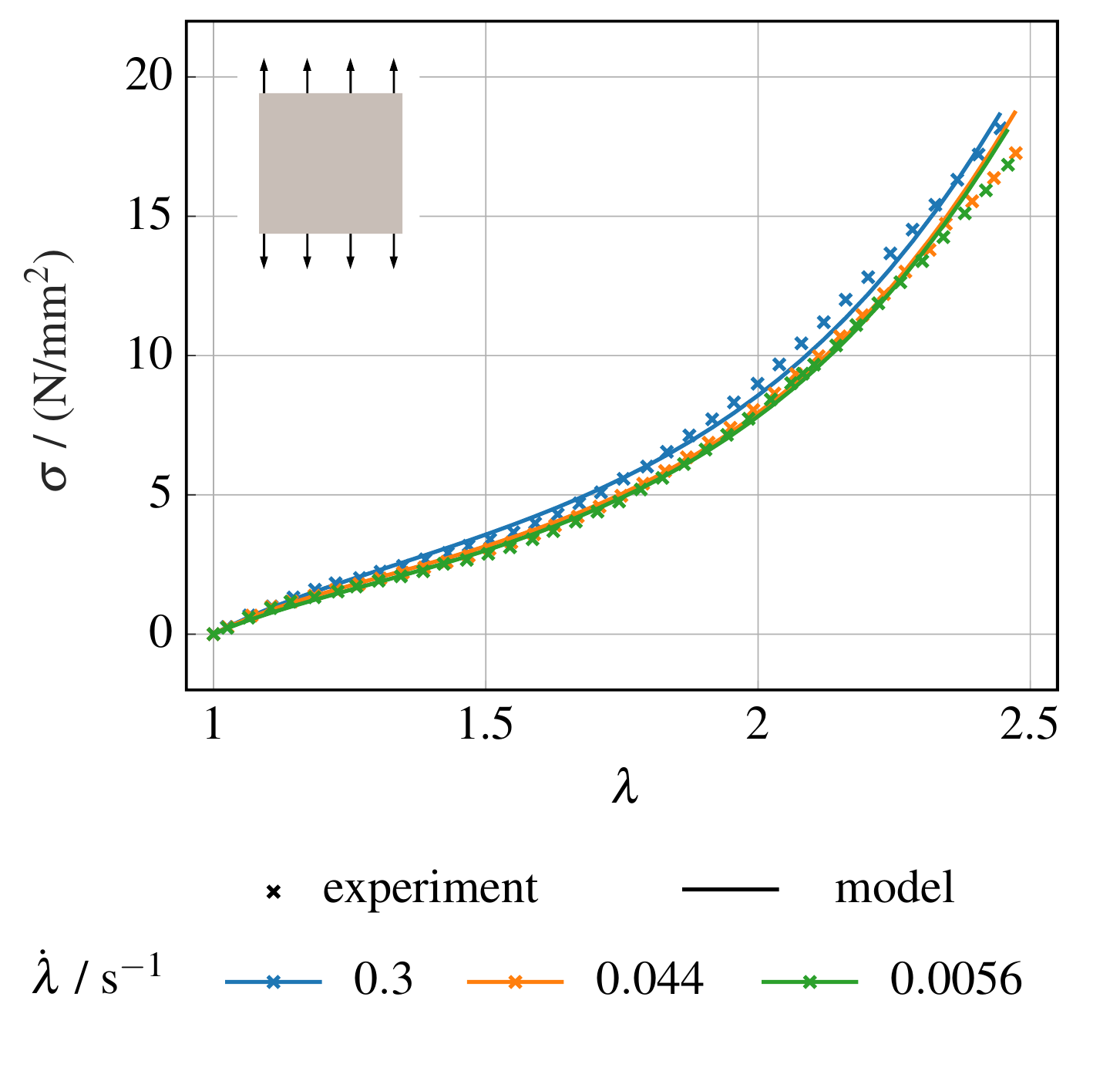} % [trim=left bottom right top]
		\caption{Stress response of the bulk material under homogeneous uniaxial tension\textemdash Experimental data \cite{loew2019} vs. present model for three different stretch rates $\dot \lambda$}
		\label{fig:fit_uniax}
\end{figure}

Within this publication, the viscoelastic behaviour of EPDM rubber is considered that has been experimentally analysed in \cite{loew2019}.  
\paragraph{Bulk response.}
At first, the parameters describing the deformation of the bulk material are determined.
For this purpose, the averaged stress-stretch curves from \cite[Fig.6]{loew2019} are considered as depicted in Fig.~\ref{fig:fit_uniax}.
For three different rates of deformation, these curves have been identified from uniaxial tension tests with dumbbell specimens.
A homogeneous uniaxial stress state is assumed and damage is not taken into account, here. Furthermore, as no information on deformation in transversal direction is available, $\nueqn=\nuovn=0.48$ is set in order to account for the high resistance against volumetric deformation that is typically observed for rubber.
For isochoric and volumetric deformation, an identical relaxation time
\begin{equation}
\tau = \frac{\etai}{2 \, \muovn} = \frac{\etav}{\kov}
\end{equation} 
is assumed.
The Ogden parameters $\mueq, \aleq, \muov, \alov$ as well as $\tau$ are then identified by means of minimising the deviation between experimental data and model prediction.
In doing so, following \cite[p. 305]{ogden1997}, it is demanded that the constants satisfy the requirements
\begin{equation}
 \aleq \, \mueq \geq 0 \glmand \aleq \in (-\infty,-1) \cup (2,\infty)
\end{equation}
for any $p$ and similar constraints for the non-equilibrium branch.
To this end, in Matlab R2020b, the \textit{GlobalSearch} strategy together with the \textit{fmincon} algorithm for constrained optimisation problems is employed.%
\footnote{A proof of uniqueness of the parameters identified, i.e. a global minimum of discrepancy between model and experiment, can not be provided. Nevertheless, \textit{GlobalSearch} involves minimisation for a huge number of different start values in order to obviate local minima.}
For an adequate approximation of the material behaviour, two Ogden exponents have revealed necessary for both the equilibrium and over-stress branch, respectively, i.e. $\Neq=\Nov=2$.%
\footnote{An increase of the number of Ogden branches to $\Neq=\Nov=3$ did not lead to a perceptibly better approximation.}
The parameters obtained are summarised in Tab.~\ref{tab:matpar-bulk}. From Fig.~\ref{fig:fit_uniax} it becomes clear that the finite viscoelasticity formulation together with the Ogden approach allows for a very good approximation of the experimental results over the entire range of stretch $\lambda \in [1,2.5]$ that has been experimentally investigated.
Furthermore, the present model enables to capture the rate-dependent response in a more reliable manner then the linear viscoelasticity model based on the Yeoh-type strain energy density \cite{loew2019}.%
\footnote{It has to be noted that the rate-dependency perceptible in Fig.~\ref{fig:fit_uniax} is not too pronounced. Accordingly, it could be worth investigating a broader range of stretch rates, since the rate-dependent behaviour of EPDM rubber can play a crucial role when it comes to failure, e.g. in case of creep fracture.
Furthermore, additional experiments such as relaxation or creep tests could allow for differentiating between equilibrium and non-equilibrium contributions to stress in a significantly more accurate manner.
However, within this contribution, we proceed with the experimental results available in the literature.}
\begin{table}[tb]
    \caption{Parameters of the finite viscoelasticity model for the deformation of the bulk material}
    \centering
    \footnotesize
    \resizebox{\linewidth}{!}{
    \begin{tabular}{c c c c c c c c c}
    \hline
    \\
        $\nueqn$ & $\mueq[1] / (\nv{N/mm})^2$ & $\aleq[1]$&$\mueq[2] / (\nv{N/mm})^2$ & $\aleq[2]$ \\
        0.48 & -1.103   &     -4.883   &  0.0105 &   7.951\\
        \\
        $\nuovn$ & $\muov[1] / (\nv{N/mm})^2$ & $\alov[1]$&$\muov[2] / (\nv{N/mm})^2$ & $\alov[2]$ & $\tau / \nv{s}$ \\
        0.48 &  -0.385     &   -4.29 &   $10^{-6}$  &  8.4 & 2.3\\
        \hline
    \end{tabular}}
    \label{tab:matpar-bulk}
\end{table}

\begin{figure}[b]
		\centering 
		\includegraphics[width=0.45\textwidth]{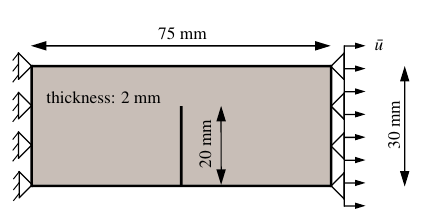} % [trim=left bottom right top]
		\caption{\textit{SENT}\textemdash Setup considered for the identification of~$\gc$}
		\label{fig:sent_setup}
\end{figure}

\paragraph{Identification of the fracture parameters.}
With the calibrated bulk deformation model at hand, the fracture phase-field is parameterized from \textit{SENT} experiments, i.e. specimens with a single pre-existing notch under tension.
These experiments have been conducted at two rates of prescribed displacement \cite{loew2019}.
The according specimen geometry is depicted in Fig.~\ref{fig:sent_setup}.

\begin{figure*}[tb!]
		\centering 
            \includegraphics[trim=0 0 0cm 0,width=0.4\textwidth]{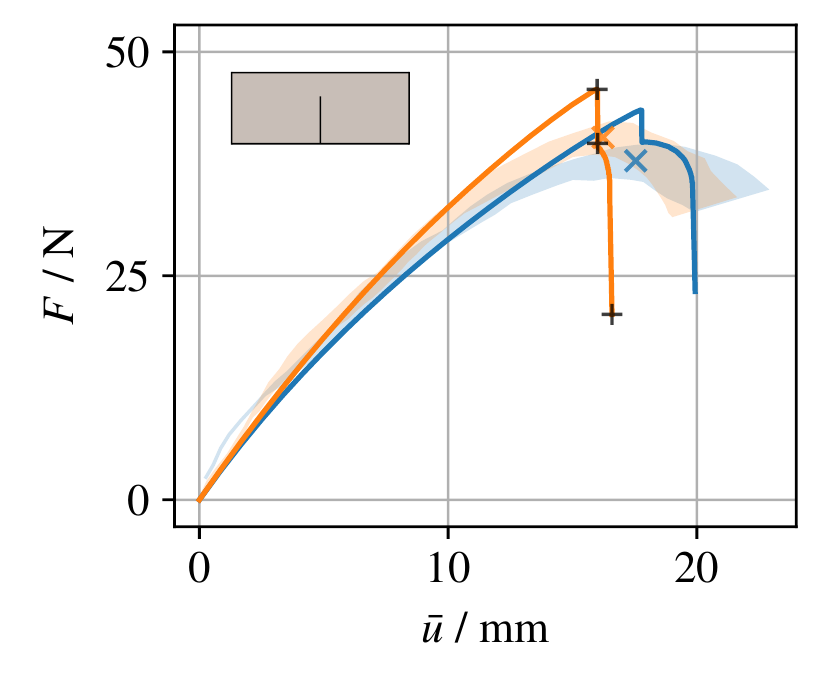} % [trim=left bottom right top]
        \hspace{5mm}
            \includegraphics[trim=0 0 0cm 0,width=0.4\textwidth]{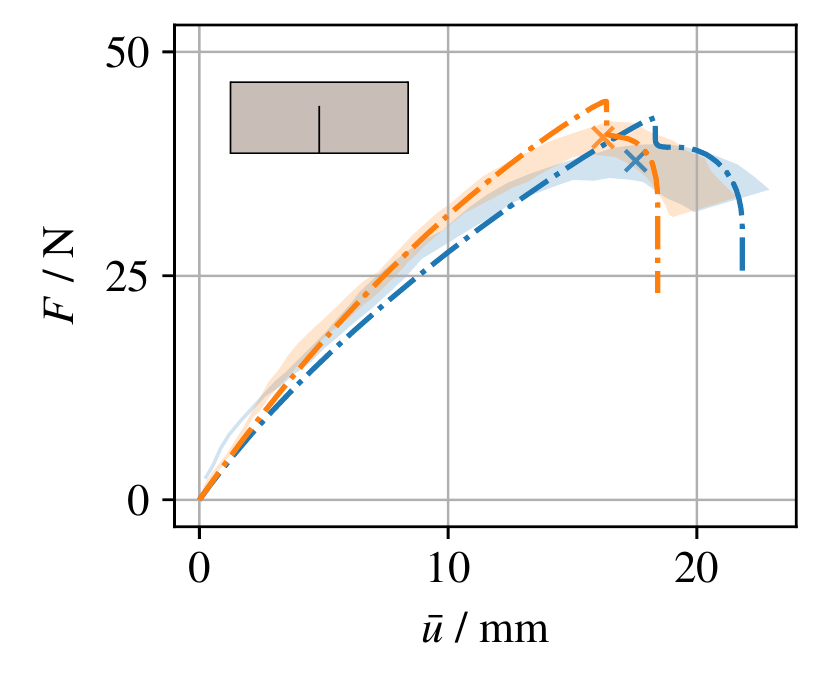} % [trim=left bottom right top]
        
        \includegraphics[trim=0 0 2cm 0,width=0.75\textwidth]{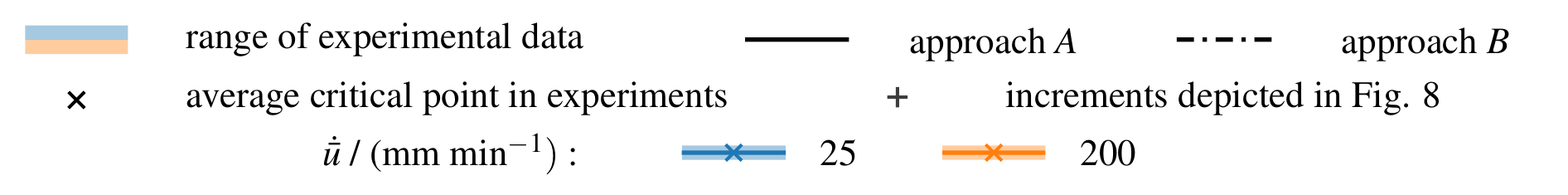} % [trim=left bottom right top]
		\caption{\textit{SENT}\textemdash experimental data~\cite{loew2019} vs. model for approaches \textit{A} ($\bvi=0$) and \textit{B} ($\bvi=1$)}
		\label{fig:fit_SENT}
\end{figure*}

\begin{figure}[tb!]
		\centering 
            \includegraphics[trim=8mm 0 8mm 0,width=0.48\textwidth]{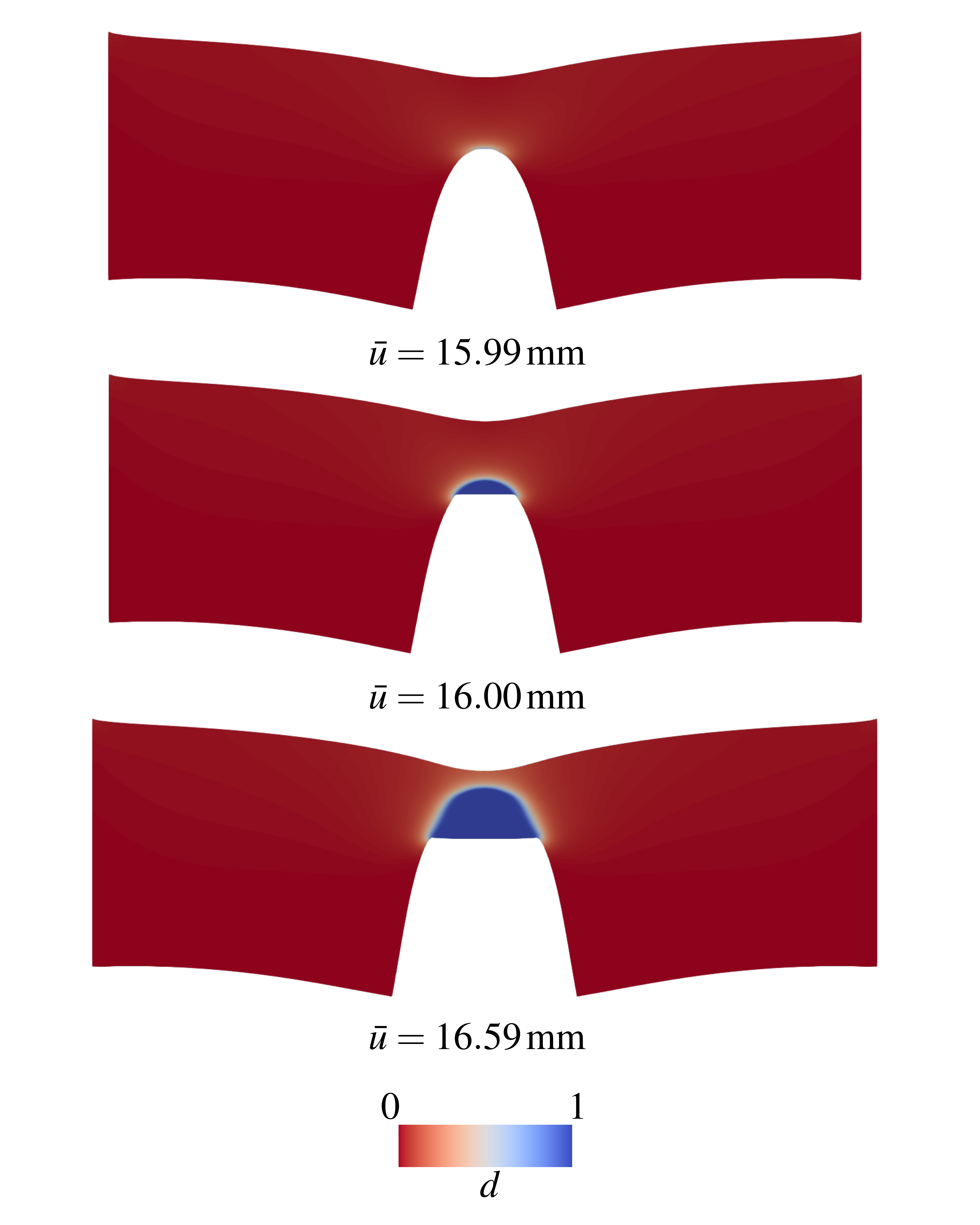} % [trim=left bottom right top]
		\caption{\textit{SENT}\textemdash crack propagation through the specimen for $\dot{\bar u}=\SI{200}{mm\per\min}$ and $\bvi=0$ (approach \textit{A}).
		The corresponding force-displacement curve is depicted in Fig.~\ref{fig:fit_SENT}.
		Qualitatively similar results are obtained for approach~\textit{B} and other $\dot{\bar u}$.}
		\label{fig:SENT_res}
\end{figure}

For the numerically motivated kinetic fracture parameter and the residual stiffness, the values $\etaf=\SI{e-4}{Ns\per\milli\meter^2}$ and $k=10^{-10}$, respectively, are chosen. In a convergence study, these values have revealed sufficiently small so that the influence of $\etaf$ and $k$ on the simulation results vanishes.
The regularisation parameter is set to $\lc=\SI{0.275}{\milli\meter}$, which is identical to \cite{loew2019}. 
In order to enable a step-by-step analysis of the model, a constant fracture toughness is assumed, here, and $\gc(\te d)$ according to \eqref{eq:rabGc-def} is investigated in Sect.~\ref{sec:rabGc}.
Furthermore, with the aim of performing a thorough analysis of viscous fracture driving force contribution in Sect.~\ref{sec:valid}, the two limiting cases $\bvi = 0$ (approach\textit{ A}) and $\bvi=1$ (approach\textit{ B}) are considered.
Under these two assumptions, the respective values of $\gc$ are identified from experimental data.
For this purpose, regarding the critical deformation in \textit{SENT} for the two rates experimentally investigated, deviation between simulation and mean values from the experiments is minimised by means of a gradient-free approach.
Since the specimens are of low thickness, plane stress conditions are assumed and two-dimensional simulations are performed, here.
Due to symmetry, only one half of the \textit{SENT} specimen is considered. The mesh consists of quadratic triangular elements and is refined along the crack path.
$h$-convergence is verified.
The optimal simulation results are compared to the range of experimental data in Fig.~\ref{fig:fit_SENT} and the parameters of the fracture model are summarised in Tab.~\ref{tab:matpar-pf-gcst}.
For both $\bvi=0$ with optimal $\gc = \SI{10.7}{N\per\milli\meter}$, and $\bvi=1$ with optimal $\gc = \SI{12.0}{N\per\milli\meter}$, good agreement between simulation and experiment can be stated.
With $\bvi=1$, a marginally better approximation is obtained for this setup.
However, in both cases, the critical force is slightly overestimated.
Furthermore, especially for the higher rate $\dot{\bar u} = \SI{3.328}{\milli\meter/\second}$, the simulated $F$-$u$ curves do not completely reproduce the smooth decrease experimentally observed in the post-critical stage preceding complete failure.
Instead, the critical point is followed by a sudden drop of reaction force that, interestingly, does not come along with complete failure yet.
It corresponds to crack initiation at the tip of the pre-existing notch, see Fig.~\ref{fig:SENT_res}, and is succeeded by a smoother decrease of force for which crack propagation through the specimen involves a slight increase of external load before, finally, it comes to complete failure.%
\footnote{A straightforward way for tuning the model such that it would better reproduce this specific experimental observation could be defining a finite $\etaf \napprox 0$, which leads to a smooth decrease of post-critical $F$-$u$ curve instead of a sudden jump, see e.g. \cite[Fig.~9]{miehe2014}. However, as outlined in Sect.~\ref{sec:pf-evol}, this approach has some important drawbacks which is why it is not pursued here.
For a more expressive investigation, it may be eligible to thoroughly elaborate on crack initiation mechanisms. For example, cavitation or void formation are often observed in rubbery polymers, see e.g. \cite{euchler2020}, and modified fracture phase-field models that explicitly take these mechanisms into account have recently been proposed in \cite{kumar2020a,kumar2021}, wherein hyperelasticity is assumed for the bulk.}
To the best of the author's knowledge, such a phenomenon does not arise in hyperelastic models, whereas it also has been reported for linear viscoelasticity \cite{loew2019,dammass2021b}.
The effect is the more pronounced the lower $\dot{\bar u}$.
Obviously, it is provoked by the rate-dependent behaviour of the bulk material that involves an increase of effective stiffness as well as the effective load bearing capacity of the material when, locally in the vicinity of the crack, rate of deformation suddenly raises up due to the initiation of fracture. 
For a rigorous analysis within the small strain context, the reader is referred to the previous work \cite{dammass2021b}.

\begin{table}[b]
    \caption{Parameters of the phase-field model calibrated for EPDM rubber with $\gc = \nv{const.}$ assumed}
    \centering
    \footnotesize

        \begin{tabular}{c c c c c c c c c}
    \hline
    \\
            & $ \etaf / (\nv{N s}/\nv{mm}^2)$  & $k$ & $\lc / \nv{mm}$ & $\bvi$ & $\gc / (\nv{N/mm})$ \\
    \\
    approach \textit{A} &  & & & 0 & $10.7$ \\
            & $10^{-4}$ & $10^{-10}$ & $0.275$ \\
    approach \textit{B} &  & & & 1 & $12.0$ \\
% %     \\
        \hline
    \end{tabular}
    \label{tab:matpar-pf-gcst}
\end{table}

\subsection{Model validation and analysis of viscous driving force}
\label{sec:valid}

\begin{figure}[tb]
		\centering 
		\includegraphics[width=0.45\textwidth]{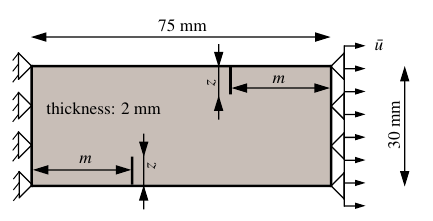} % [trim=left bottom right top]
		\caption{\textit{DENT}\textemdash Setup for model validation and analysis. For comparison of model prediction with experimental data from \cite{loew2019}, symmetrical specimens are considered, i.e. $m=\SI{75/2}{mm}$.}
		\label{fig:dent_setup}
\end{figure}

For further model validation and analysis, double notched specimens under tension (\textit{DENT}) with varying length of the pre-existing notch $z$ are considered as depicted in Fig.~\ref{fig:dent_setup}.

\begin{figure*}[tb!]
		\centering 
            \includegraphics[trim=0 0 0mm 0,width=0.85\textwidth]{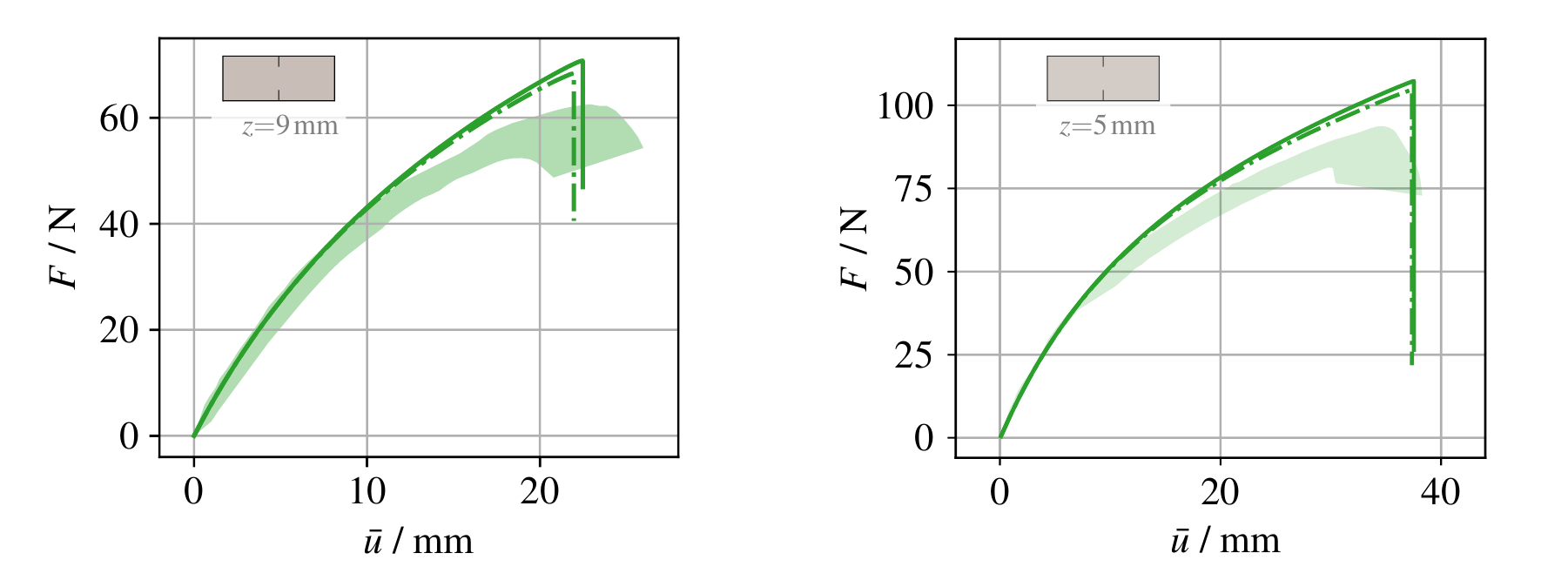} % [trim=left bottom right top]
            
                    \includegraphics[trim=0 0 1cm 0,width=0.8\textwidth]{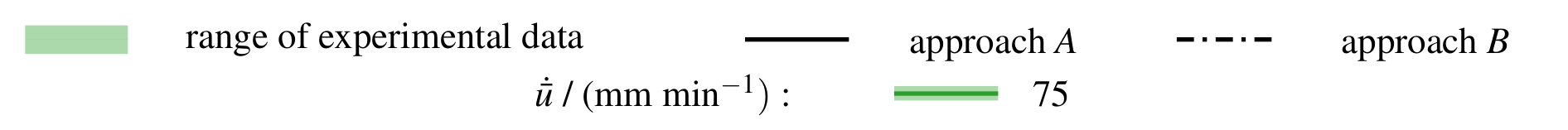} % [trim=left bottom right top]
		\caption{Symmetrical \textit{DENT}\textemdash experimental data~\cite{loew2019} vs. model prediction for two values of length of pre-existing notch $z \in \{9,5\} \, \nv{mm}$ and constant $\dot{\bar u}$. Similar results are obtained for $z= \SI{7}{mm}$, see Fig.~\ref{fig:DENT_z7}, and $z=\SI{3}{mm}$ (not depicted).}
		\label{fig:fit_DENT_zs}
\end{figure*}

At first, a symmetrical specimen geometry is considered, i.e. $m=\SI{75/2}{mm}$. The predictions of the model parameterized in the previous Sect. are compared to experimental data from \cite{loew2019} for $ z \in \{9, 5\} \, \nv{mm}$ and a constant rate $\dot{\bar u}=\SI{75}{mm\per\min}$ in Fig.~\ref{fig:fit_DENT_zs}.
For both approaches \textit{A} and \textit{B}, model predictions fit the experimental results well, which is also true for $ z \in \{7, 3\} \, \nv{mm}$ (not depicted).
The good agreement demonstrates the predictive capability of the present model and the suitability of the parameter identification from experiments with homogeneous and single-notched specimens.

\begin{figure*}[tb!]
		\centering 
            \includegraphics[trim=5mm 0 0mm 0,width=0.85\textwidth]{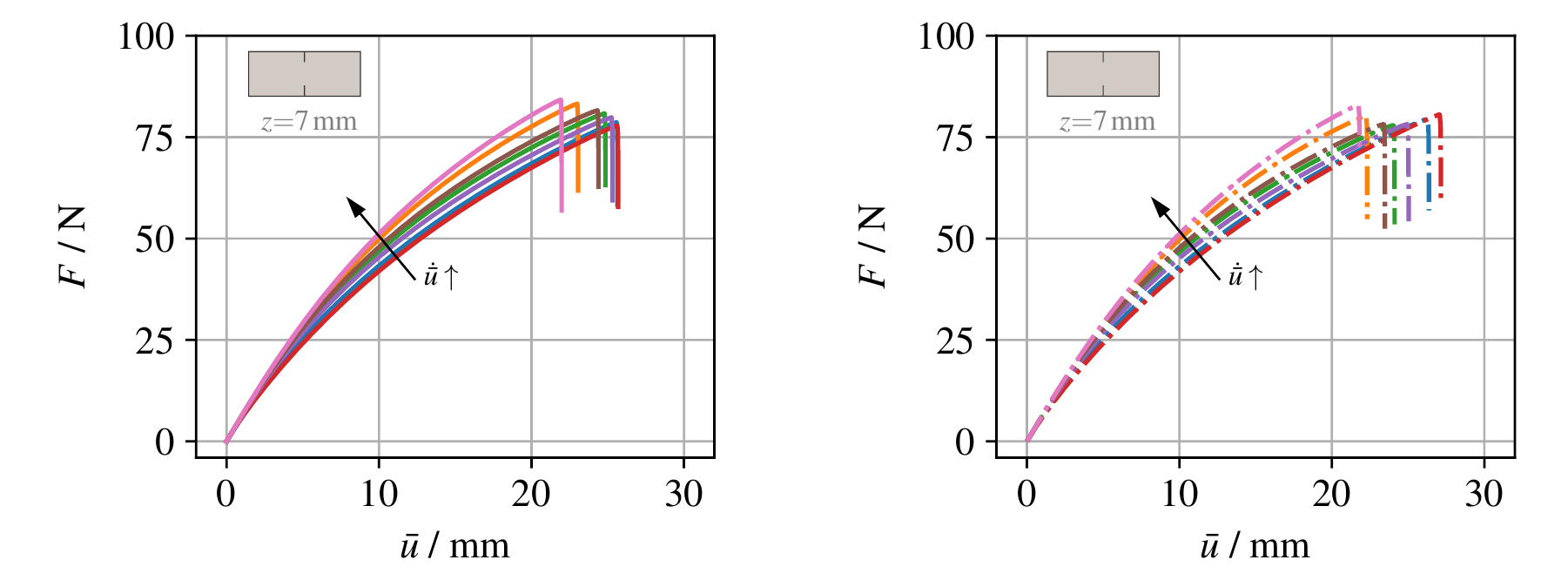} % [trim=left bottom right top]
            
            \includegraphics[trim=0 0 5mm 0,width=0.85\textwidth]{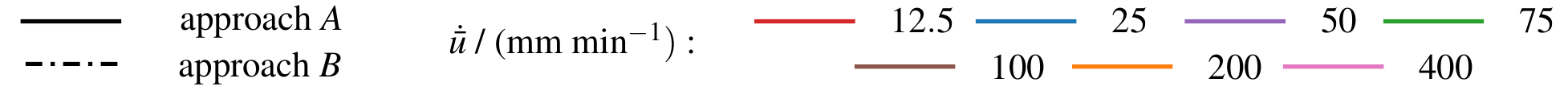} % [trim=left bottom right top]
		\caption{Symmetrical \textit{DENT}\textemdash comparison of model prediction for approaches \textit{A} and \textit{B} for various rates and a fixed size of pre-existing notch $z= \SI{7}{mm}$}
		\label{fig:DENT_z7}
\end{figure*}

\begin{figure}[tb!]
		\centering 
            \includegraphics[trim=5mm 0 0mm 0,width=0.4\textwidth]{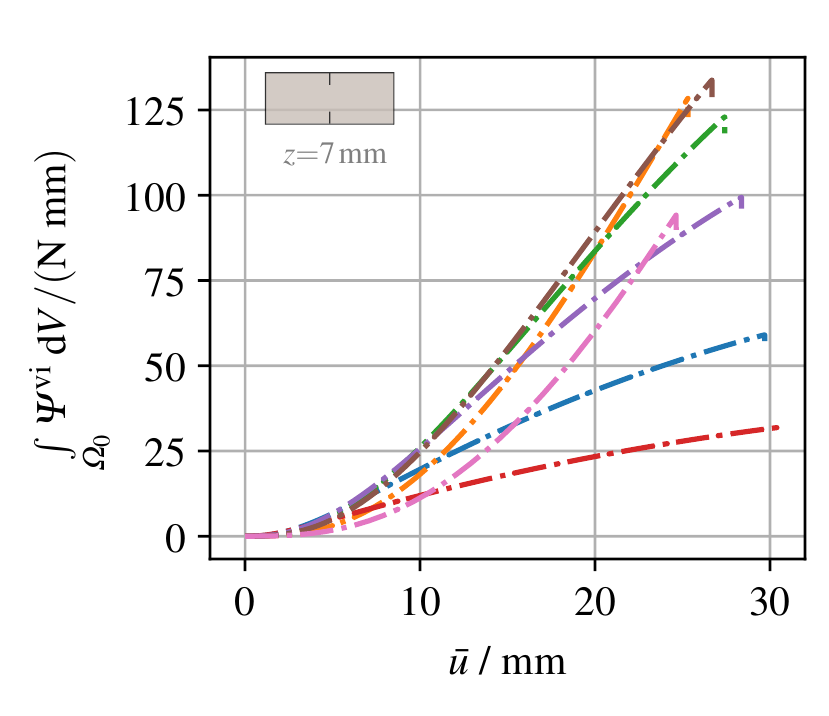} % [trim=left bottom right top]
            \includegraphics[trim=0mm 0 5mm 0,scale=0.8]{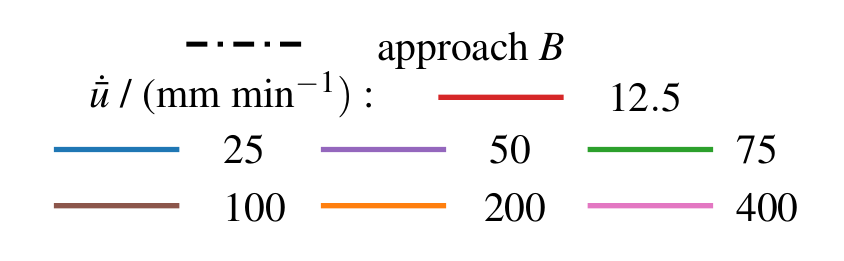}
            \caption{Symmetrical \textit{DENT}\textemdash free energy contribution related to viscous dissipation (approach \textit{B}) for various rates and a fixed size of pre-existing notch $z= \SI{7}{mm}$}
		\label{fig:DENT_psivi}
\end{figure}

With the aim of more thoroughly analysing the rate-dependency of responses and elaborating on the driving force contributions, additional simulations are performed for $z= \SI{7}{mm}$ and various rates $\dot{\bar u} \in [12.5, 400] \, \si{mm\per\min}$.
The numerical predictions for the two approaches $\bvi=0$ and $\bvi=1$ are compared in Fig.~\ref{fig:DENT_z7}.
Regardless of the approach for the driving force, for high displacement rates, the responses converge against an upper elastic limit for which there is almost no viscous dissipation until failure. For very low $\dot{\bar u} $, the responses of the structure likewise approach a lower elastic limiting case where over-stresses do approximately vanish during entire simulation.
In between, for intermediate displacement rates, the critical displacement level diminishes with $\dot{\bar u}$ for both approaches \textit{A} and \textit{B}.
In contrast, regarding the rate-dependency of critical force level, the model predictions do significantly differ depending on whether a viscous fracture driving force contribution is assumed or not.
On the one hand, critical force monotonically increases with rate when there is no such contribution, i.e. $\bvi=0$ \textit{(A)}. On the other hand, for $\bvi=1$ \textit{(B)}, critical force becomes minimal for intermediate $\dot{\bar u}$, for which the greatest critical values of $\Psivi$ are observed, see Fig.~\ref{fig:DENT_psivi}.
Although no experimentally-determined force-displacement curves are available, it can be stated that the former is in agreement with experimental observations~\cite{loew2019}, whereas the latter contradicts experimental experience.
At least when modelling fracture of elastomeric materials under monotonic loading, in some cases,  fracture driving force contribution associated to accumulated viscous dissipation can thus lead to erroneous model predictions.
In other words, modelling approach \textit{A} has revealed more plausible, which, in a sense, is different from plasticity, where a fracture driving force related to inelastic mechanisms has revealed advantageous \cite{ambati2015,borden2016}.
Interestingly, such an observation has not been made in the previous study within the small strain framework~\cite{dammass2021b}, where a less pronounced influence of viscous effects on crack propagation has been observed. This can probably be attributed to the fact that the present formulation enables to describe larger deviations away from thermodynamic equilibrium, resulting in considerably greater viscous contributions to fracture driving force.

As it has been comprehensively described in \cite{dammass2021b}, it essentially is the change of effective stiffness and the amount of dissipation until failure that lead to the change of critical force and displacement level with rate of external load.
While the amount of fracture driving force necessary for crack growth remains constant, the fracture driving force available for a constant level of deformation can change with rate.
On the one hand, effective stiffness of the viscoelastic material monotonically increases with increasing rate of deformation. For a certain external displacement $\bar u$ prescribed, the density of strain energy raises with $\dot{\bar u}$, accordingly.
On the other hand, in case of monotonic loads, the amount of viscous dissipation and thus, in case of $\bvi>0$, the level of $\Psivi$ at failure becomes maximal for intermediate rates. 

Although viscous fracture driving force contribution has revealed not suitable for describing failure of elastomers under monotonic loads, it might be suitable for other classes of materials, e.g. thermoplastics, and especially for the modelling of fatigue fracture, e.g. with $0 < \bvi \ll 1$.
In composites and thermoplastic materials, for instance, viscous dissipation and self-heating mechanisms can have an important influence on fatigue life, cf.~\cite{mortazavian2015}.%
\footnote{For example, in the phase-field fatigue fracture model \cite{loew2020b}, which is applied to a rubbery polymer, a fatigue fracture driving force is introduced that also incorporates viscous dissipation.
However, similar to \cite{loew2019}, a model of linear viscoelasticity at finite deformation is used which does not allow for separation of accumulated viscous dissipation and non-equilibrium part of stored strain energy.
As a consequence, entire viscous dissipation is also included in the quasi-static fracture driving force contribution.}

\paragraph{Crack patterns in asymmetrical specimens.}
\begin{figure*}[tbp!]
		\centering 
            \includegraphics[trim=0mm 0 15mm 0,width=0.4\textwidth]{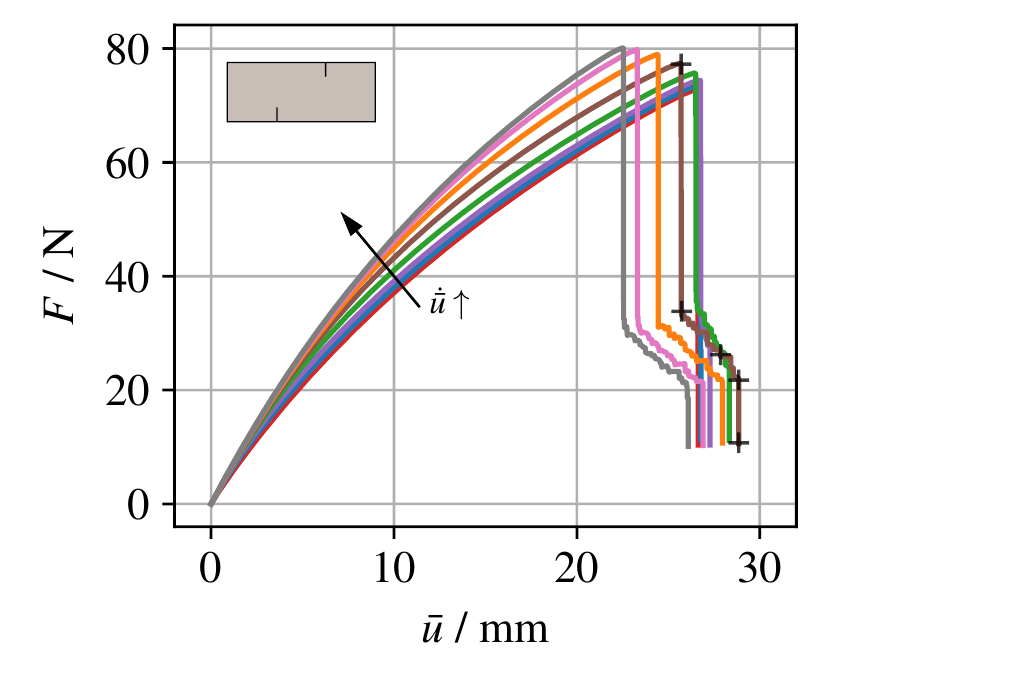} % [trim=left bottom right top]
            \includegraphics[trim=0mm 0 15mm 0,width=0.4\textwidth]{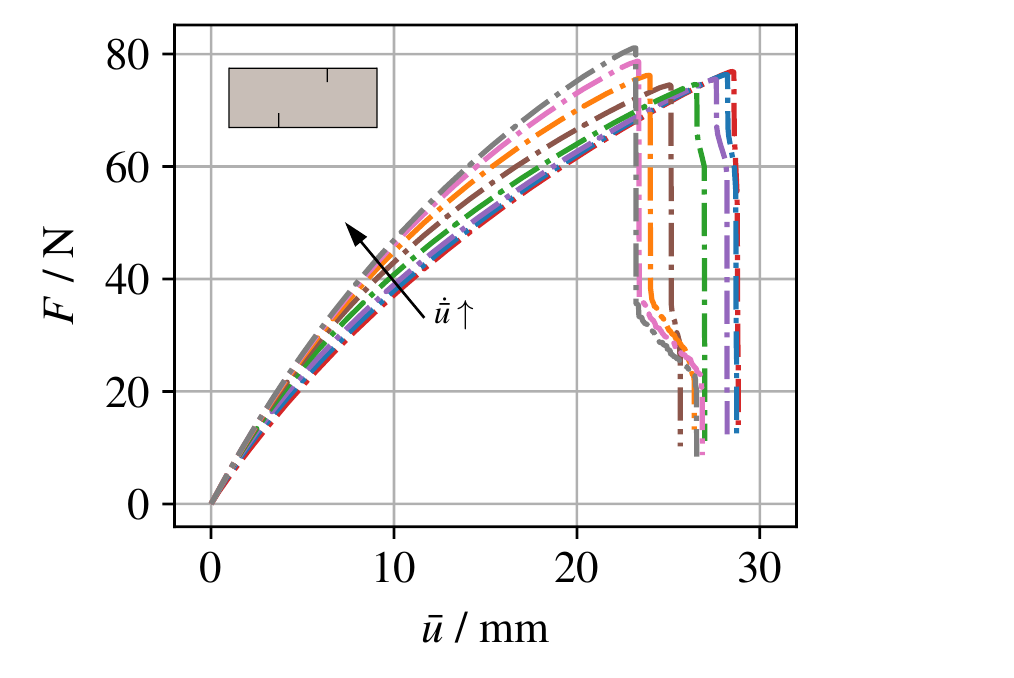} % [trim=left bottom right top]
            \includegraphics[trim=0 0 5mm 0,width=0.67\textwidth]{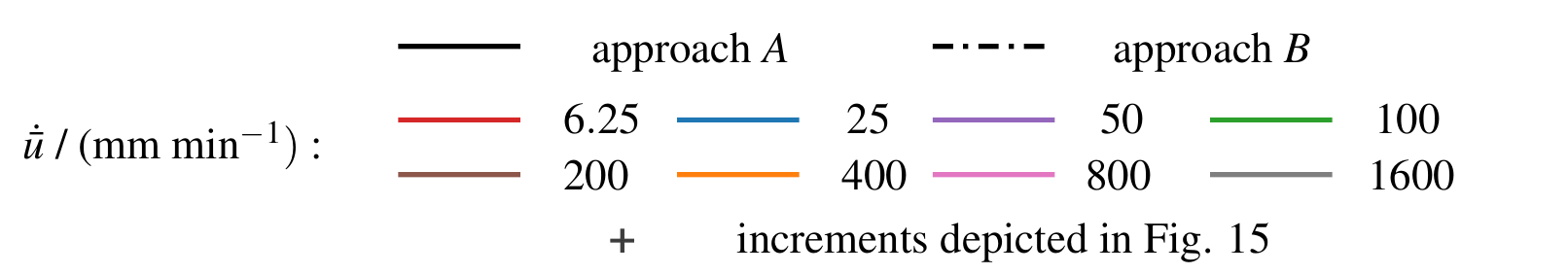} % [trim=left bottom right top]
		\caption{Asymmetrical \textit{DENT}\textemdash comparison of specimen responses for approaches \textit{A} and \textit{B} for various rates, pre-notch position $m=\SI{27.5}{mm}$ and notch length $z= \SI{7}{mm}$  }
		\label{fig:DENT_asymm_fu}
\end{figure*}
\begin{figure*}[tbp!]
		\centering 
    \includegraphics[trim=0mm 0 0mm 0,width=0.9\textwidth]{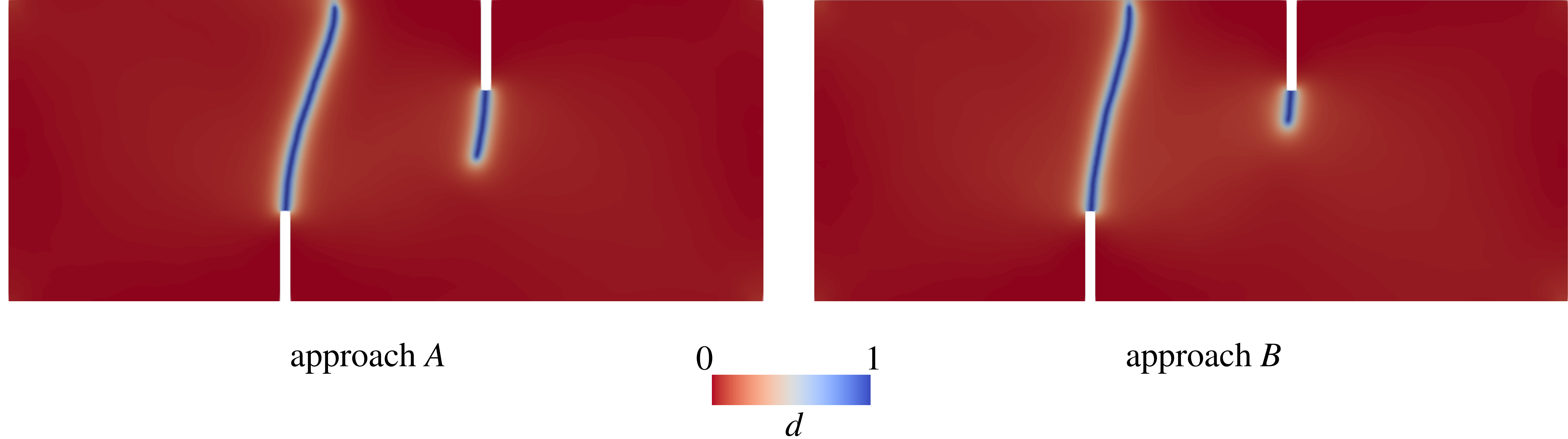}
		\caption{Asymmetrical \textit{DENT}\textemdash final crack patterns in the reference configuration $\omref$ for approaches \textit{A} and \textit{B} and $\dot{\bar{u}} = \SI{200}{mm\per\min}$. 
		For the EPDM rubber considered, no experimental results are available for this setup, yet the crack paths resemble experimental observations made for other viscoelastic materials, see e.g. \cite{han2012}.}
		\label{fig:DENT_asymm_rissRef}
\end{figure*}
\begin{figure}[tbp!]
		\centering 
            \includegraphics[trim=0 0 0cm 0,width=0.45\textwidth]{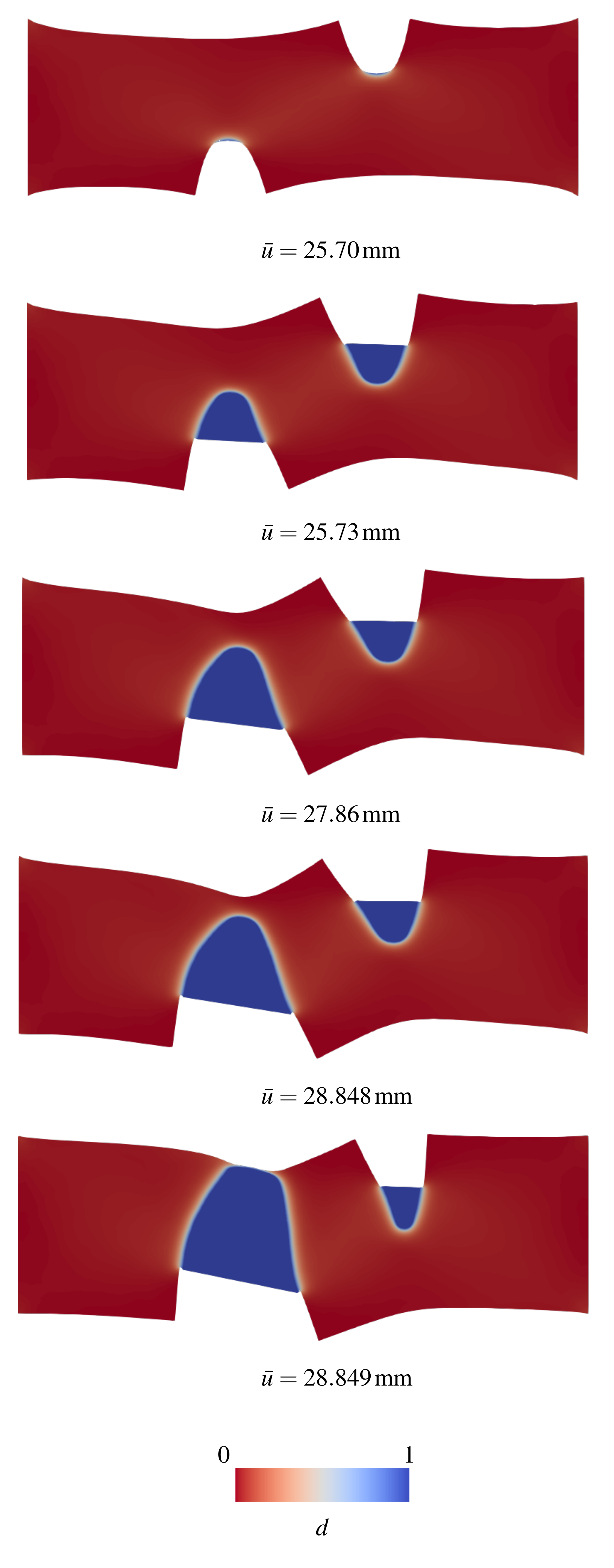} % [trim=left bottom right top]
		\caption{Asymmetrical \textit{DENT}\textemdash phase-field crack initiation and propagation through the specimen for $\dot{\bar u}=\SI{200}{mm\per\min}$ and $\bvi=0$ (approach \textit{A}).
		The corresponding force-displacement curve is depicted in Fig.~\ref{fig:DENT_asymm_fu}.
		Qualitatively similar results are obtained for approach~\textit{B} and other $\dot{\bar u}$.}
		\label{fig:DENT_res_asymm}
\end{figure}
In addition to the symmetrical specimens, simulation results are presented in the following for an asymmetrical \textit{DENT} geometry as depicted in Fig.~\ref{fig:dent_setup} with $m=\SI{27.5}{mm}$ and $z=\SI{9}{mm}$.
Since for ductile fracture of metals, where instead of viscoelasticity another class of dissipative materials is involved, the choice of fracture driving force revealed crucial the appropriate numerical description of asymmetrical crack patterns, cf. \cite{ambati2015}, simulations are performed for both approaches \textit{A} and \textit{B}.
The corresponding force-displacement curves are depicted in Fig.~\ref{fig:DENT_asymm_fu}. The overall rate-dependency of the specimen response is identical to what has been described above for the symmetrical geometry. In particular, for $\bvi=1$, the numerically predicted critical force becomes minimal for an intermediate rate of external displacement, which does hardly coincide with what would be observed in experiments.
In Fig.~\ref{fig:DENT_asymm_rissRef}, the final crack patterns are compared for $\dot{\bar u}=\SI{200}{mm\per\min}$. In order to ease comparison, the phase-field is shown with respect to the reference domain $\omref$.
For both fracture driving forces \textit{A} and \textit{B}, the crack pattern predicted for the viscoelastic material is essential different from what is typically observed when metals fail in a ductile manner.
Instead of a single crack that connects the two pre-existing notches, two cracks independently propagate through the specimen.
At a certain length, one of the two stops to propagate, resulting in an asymmetrical final crack pattern, see Fig.~\ref{fig:DENT_res_asymm}.
Regardless of $\bvi$ and $\dot{\bar u}$, qualitatively identical crack paths are predicted.%
\footnote{For all the simulations performed, it is always the right crack tip which stops propagating at a certain length. It is deemed likely that this is due to the non-symmetric mesh that has been used for all the computations.}
However, depending on $\bvi$, slight differences concerning the final length of the shorter crack can be stated especially for intermediate rates.
Interestingly, when critical force is reached, the two cracks suddenly propagate over a finite width, which comes along with a significant abrupt drop of force. For intermediate and higher rates, similar to \textit{SENT} geometry, a slight increase of external displacement $\bar u$ is necessary to make one of the cracks propagate further, resulting in a less heavy slope of the force-displacement curve before it finally comes to catastrophic failure.
For these higher rates, in the simulations there is a stage that can be seen as a kind of \textit{stick-slip}-like  crack propagation, where the crack tip suddenly advances over a finite distance and then arrests over and over again. These effects also lead to a non-\textit{smooth} $F$-$u$ curve in the post-critical range.
Interestingly, for very small $\dot{\bar u}$, such a behaviour is not simulated.
In the literature on dynamic crack growth, comparable phenomena have been reported, cf. e.g. \cite{hageman2021}.
However, it has to be noted that regarding this particular aspect, the predictive capabilities of the present model are somewhat limited, as inertia effects are not taken into account.

For the EPDM rubber for which the model has been parameterized here, no experimental results are available for crack propagation in asymmetrical specimens.
Nevertheless, the crack patterns simulated with the present model are in excellent agreement with what has been observed in experiments for other viscoelastic materials, see e.g. \cite{han2012}.
It is obvious that, when specimen geometries are similar, these crack patterns in viscoelastic materials can differ from the ones that form in elasto-plastic ones, since the inelastic mechanisms are essentially different. For example, there typically is no zone of inelastic localisation in viscoelastic materials whereas localisation of plastic deformation can play an important role when it comes to ductile fracture of metals.

\begin{figure*}[tb!]
		\centering 
            \includegraphics[trim=10mm 0 0mm 0,scale=0.85]{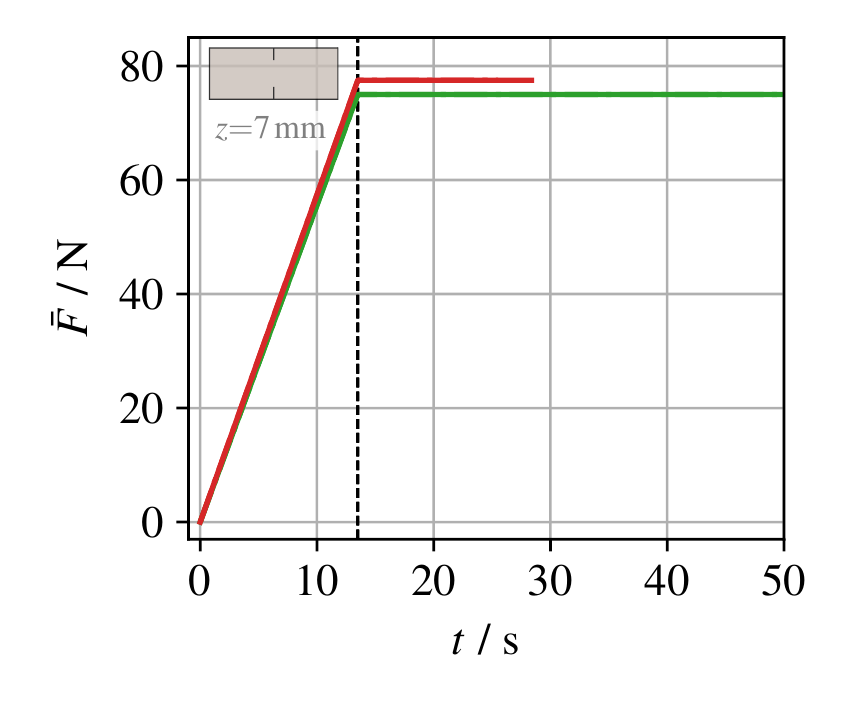} % [trim=left bottom right top]
            \hspace{10mm}
            \includegraphics[trim=0 0 5mm 0,scale=0.85]{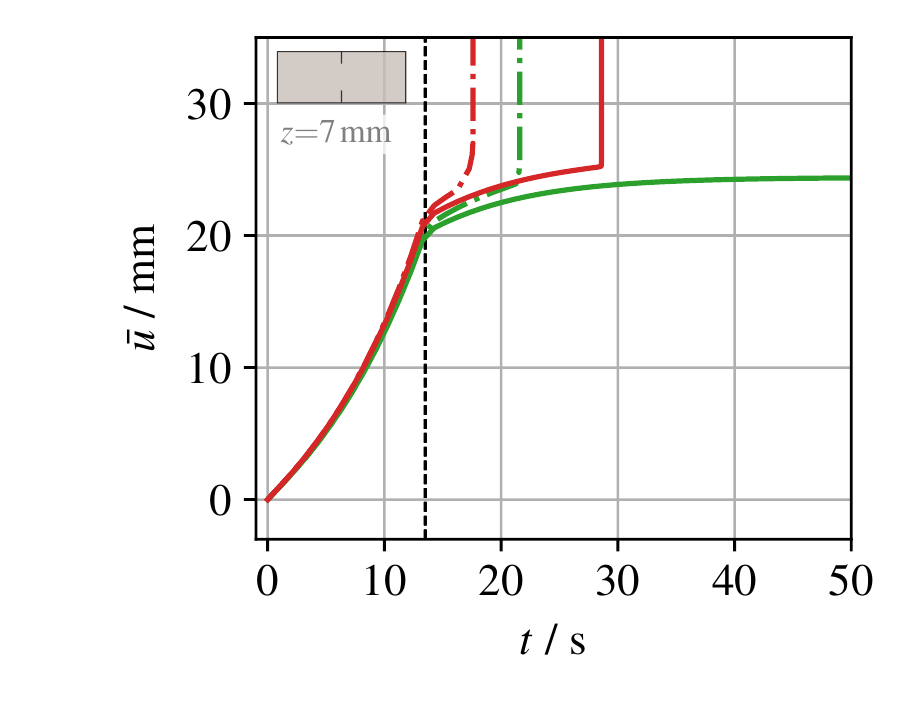} % [trim=left bottom right top]
            \includegraphics[trim=0 0 5mm 0,scale=0.7]{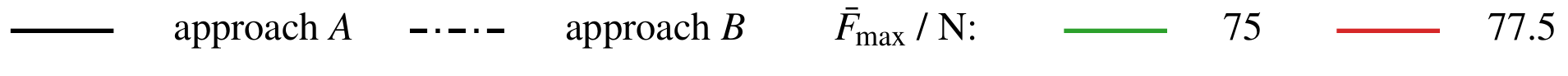} % [trim=left bottom right top]
		\caption{\textit{DENT}\textemdash boundary conditions and model predictions for the investigation of creep fracture}
		\label{fig:DENT_Creep}
\end{figure*}
\paragraph{Creep fracture.}
In addition to fracture under monotonically increasing loads, a qualitative analysis of creep fracture is performed by means of one representative example.
For this purpose, the symmetrical \textit{DENT} geometry with $z=\SI{7}{mm}$ is revisited. Instead of $\bar u$, a traction force $\overline{F}$ is applied that linearly increases with time until a certain value $\overline{F}_\nv{max}$ is reached and is hold constant, subsequently.
For two different values of $\overline{F}_\nv{max}$, boundary conditions and model predictions are depicted in Fig.~\ref{fig:DENT_Creep} for both approaches \textit{A} and \textit{B}.
It can be stated that, generally, creep fracture can be captured regardless of the value of the assumption made on fracture driving force.%
\footnote{For the specific setup considered here, no experimental results are available. Nevertheless, from \cite{loew2019}, it can be reasoned that for both the lower and the higher value of $\overline{F}_\nv{max}$ considered here, creep fracture would have to be expected in an experiment which is not captured in case of approach \textit{A}. 
However, this deviation is assumed to essentially arise from the lack of non-monotonic experimental data for parameterization of the viscoelastic bulk deformation model.
Since the model could solely be calibrated from monotonic experiments, an uncertainty of the prediction in case of creep loads can not be avoided.}
In case of $\bvi>0$, failure can occur for lower $\overline{F}_\nv{max}$ and after a shorter amount of creep time than for $\bvi=0$. Furthermore, if a fracture driving force contribution from viscous dissipation is assumed, it can also depend on the rate $\dot{\overline{F}}$ if creep fracture is predicted, since viscous dissipation vanishes for very small $\dot{\overline{F}}$, see \cite{dammass2021b} for a discussion in the small strain context.

\subsection{Investigation of rate-dependent fracture toughness}
\label{sec:rabGc}
In the foregoing Section and the previous work \cite{dammass2021b}, it is demonstrated that within the scope of an energetic phase-field fracture approach, a rate-dependent material model for the bulk induces a certain relationship between critical load and rate of deformation when $\gc$ is constant.
Therefore, in addition to experimental indication \cite{gent1994,gent2012,goh2005,gamonpilas2009,forte2015}, there also is a clear motivation for assuming a rate-dependent toughness from a phenomenological point of view.
Assuming $\gc$ to be a function of effective rate of deformation $r = \left\Arrowvert \te d \right\Arrowvert_\nv{F}$ enables more flexibility in describing the rate-dependent failure of varied materials.
In what follows, this is demonstrated by means of numerical studies considering both an increase and a decrease of $\gc$ with $r$.
For this purpose, the \textit{DENT} setup with $z=\SI{7}{mm}$  and $\bvi=0$ is revisited.
For $\gc$, the sigmoid-shaped function~\eqref{eq:rabGc-def} is assumed with $\gco=\SI{10.7}{N/mm}$ and $\bvi=0$ as parameterized for EPDM whereas the responses for different $\gct>\gco$ as well as $\gct < \gco$ are investigated.
Apart from that, the parameters are identical to the ones listed previously.

\begin{figure}[tb!]
		\centering 
            \includegraphics[trim=5mm 0 0mm 0,width=0.4\textwidth]{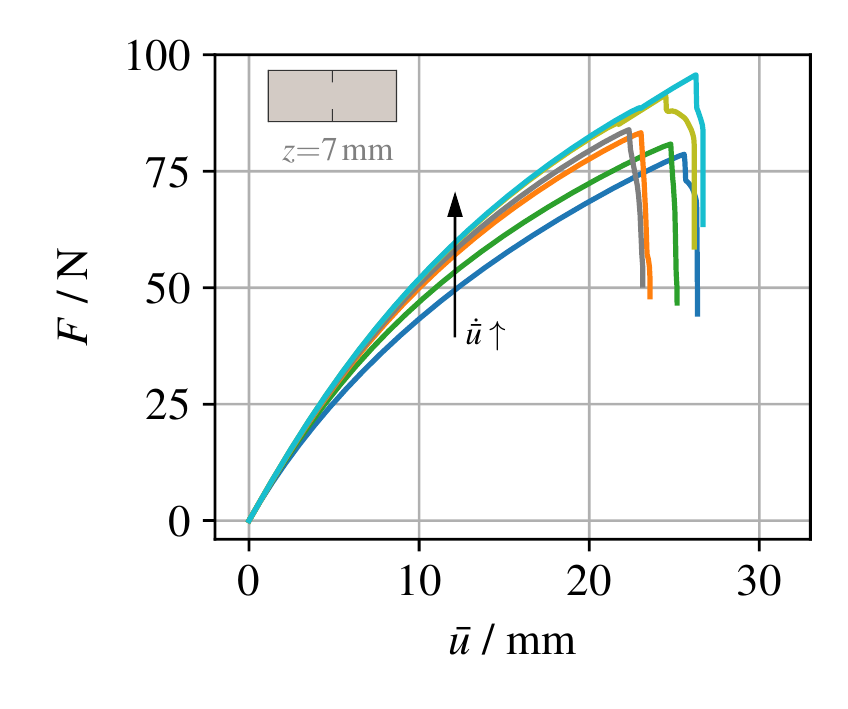} % [trim=left bottom right top]
            
            \includegraphics[trim=0 0 5mm 0,scale=0.8]{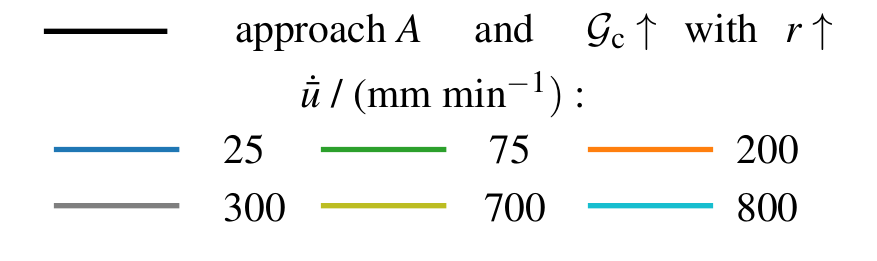} % [trim=left bottom right top]
		\caption{\textit{DENT}\textemdash model prediction in case of fracture toughness~$\gc$ assumed to increase with effective rate of deformation~$r$}
		\label{fig:DENT_GcIncr}
\end{figure}

The case of $\gc$ increasing with rate of deformation is investigated first. As a representative example, the specimen response is depicted in Fig.~\ref{fig:DENT_GcIncr} for $\gct = 2 \, \gco$, $\rref=\SI{200}{s^{-1}}$, $c=10/\rref$.
For this specific choice of $\rref$, before it comes to crack propagation, the effective rates of deformation $r$ satisfy $r \ll \rref$ within the entire domain for all $\dot{\bar u} \lesssim \SI{300}{mm/min}$.
Through comparison of Figs.~\ref{fig:DENT_GcIncr} and \ref{fig:DENT_z7}, it becomes clear that for these smaller rates, the pre-critical range of the specimen response is identical to the case where $\gc = \gco = \const $ In particular, effective stiffness and critical force raise with rate $\dot{\bar u}$, whereas critical deformation decreases. 
When the critical point is reached and crack propagation starts, effective rate of deformation $r$ suddenly raises up within the material, resulting in an increase of $\gc\left( r(\te d) \right)$. Accordingly, in the post-critical range of the $F$-$u$ curves, a slightly less sharp slope can be observed with respect to $\gc = \const$
However, this effect is not very pronounced compared to the effects arising from the rate-dependent toughness when pre-critical rate of deformation $r$ becomes close to the threshold value $\rref$.%
\footnote{It has to be noted that, when crack propagation takes place, quantitative predictive capability regarding the rate of deformation is somewhat limited for present formulation, since inertia effects are not taken into account.}
In that case, deformation at failure begins to raise with rate similar to stiffness and critical force.
Experimentally, similar effects can be observed for some natural materials, see e.g. \cite{schuldt2018} for an overview, as well as viscoelastic silicone elastomer based model systems \cite{boisly2016}.

\begin{figure}[tb!]
		\centering 
            \includegraphics[trim=5mm 0 0mm 0,width=0.4\textwidth]{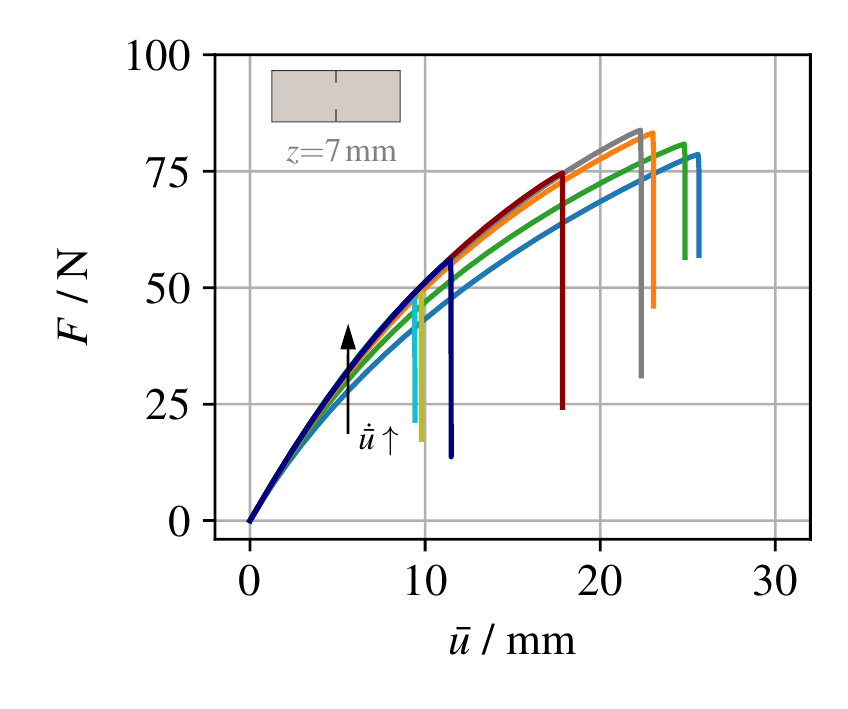} % [trim=left bottom right top]
            
            \includegraphics[trim=0 0 5mm 0,scale=0.8]{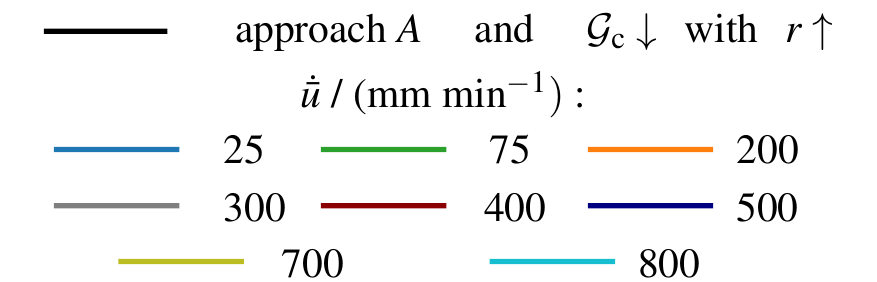} % [trim=left bottom right top]
		\caption{\textit{DENT}\textemdash model prediction in case of fracture toughness~$\gc$ assumed to decrease with effective rate of deformation~$r$}
		\label{fig:DENT_GcDecr}
\end{figure}

For the discussion of $\gc$ decreasing with $r$, $\gct=\gco/4$, $\rref=\SI{200}{s^{-1}}$, $c=10/\rref$, are considered, exemplary.
From the force-displacement curve depicted in Fig.~\ref{fig:DENT_GcDecr} it appears that for $\dot{\bar u} \gg \SI{300}{mm/min}$, the responses do again coincide with the case $\gc = \gco = \const$ Naturally, the initiation of the phase-field crack at the notch tips is immediately followed by complete failure, since in this moment, the sudden increase in rate of deformation comes along with a drop of toughness.
Nevertheless, for the \textit{DENT} geometry, similar behaviour is obtained as simulation result for $\gc=\const$, which is in agreement with experiments.
For high rates $\dot{\bar u} \geq \SI{400}{mm/min}$, where $ r \gtrsim \rref$ also holds in pre-critical range, the decrease of deformation of failure that stems from the rate-dependent stiffness of the viscoelastic material is further intensified by the rate-dependent fracture toughness.
In addition, critical force does no longer raise up with $\dot{\bar u}$ yet also decreases.
For sugar-based confections \cite{schab2021}, a similar characteristic behaviour has been observed very recently. For high displacement rates, these materials fail in a brittle manner, i.e. at small deformation as well as low external force, whereas at low rates, they can undergo large deformation.\footnote{A publication on experimental and numerical investigation of this \textit{brittle-to-ductile fracture mode transition} is in preparation.}

\section{Conclusion and outlook}
\label{sec:concl}

For the simulation of fracture of materials with rate-dependent behaviour, a flexible phase-field model is presented.
To this end, the theory of finite viscoelasticity~\cite{reese1998} is adopted for the deformation of the bulk material. The phase-field model is formulated such that, depending on the choice for the parameters, a portion of viscous dissipation can enter the fracture driving force.
Moreover, in addition to the viscoelastic model of the bulk material, a fracture toughness function that depends on rate of deformation can be considered.

In order to analyse the coupling between different rate effects, a gradual analysis of the model is performed.
The model of finite viscoelasticity is parameterized for an EPDM rubber based upon stress-deformation curves from the literature. Ogden-type strain energy densities are considered for both the equilibrium and over-stress parts of the response and very good agreement of the model with experimental data is obtained.
Assuming a constant fracture toughness for the EPDM rubber, two limiting cases are studied regarding the fracture driving force and the respective values of toughness are identified from experimentally-determined \textit{SENT} force-displacement curves. In doing so, either entire viscous dissipation or only effectively stored strain energy is assumed to enter the fracture driving force, respectively.
In the absence of a driving force contribution related to viscous dissipative mechanisms, very good agreement between model predictions and experiments can be stated for different setups. In this case, plausible results are obtained over a broad range of rates of external load and deformation, respectively.
On the contrary, if viscous dissipation is assumed to enter fracture driving force, erroneous model predictions can arise, here. In this case, agreement with experimental data is obtained for some specific rates, only. 
Accordingly, different from e.g. phase-field modelling of ductile fracture in metals, a distinct fracture driving force contribution related to inelastic dissipative mechanisms as proposed in \cite{loew2019,loew2020} or \cite{shen2019} has revealed not favourable for viscoelastic materials, in particular not for rubbery polymers.
Furthermore, comparing the crack paths predicted in asymmetrical \textit{DENT} specimens, it is demonstrated that such a driving force contribution is not necessary in order to predict non-symmetric crack patterns in an appropriate manner. 

By means of a numerical study, it is demonstrated that a rate-dependent fracture toughness can significantly increase the capability of the phase-field model in capturing varied experimentally-observable responses.
In particular, it seems suitable to describe rate-dependent brittle-to-ductile fracture mode transitions.
In contrast, in case of a constant toughness, the rate-dependent model of bulk deformation induces a certain rate-dependency of critical stress and deformation, which does not coincide with experimental evidence for some specific materials.
At least from a phenomenological point of view, rate-dependent fracture toughness thus seems to be an essential tool for modelling of rate-dependent fracture phenomena.
While this contribution clearly demonstrates the potential of a rate-dependent fracture toughness within the proposed model, a quantitative description of rate-dependent brittle-to-ductile fracture mode transitions is beyond its scope.
A thorough experimental analysis of these effects in materials with rate-dependent deformation behaviour, e.g. caramel-based confections \cite{schab2021}, as well as a quantitative description based upon the framework presented in this contribution are the subject of current work.

\section*{Acknowledgements}
Support for this research was provided by the German Research Foundation (DFG) under grant KA 3309/9-1.

The authors gratefully acknowledge Jörg Brummund for the fruitful discussions.
The computations were performed on a HPC cluster at the Centre for Information Services and High Performance Computing (ZIH) at TU Dresden. The authors thank the ZIH for allocation of computational time.

% REFERENCES
\bibliographystyle{spphys}       
{\small \bibliography{bibfile}}

\appendix
\section{Tangent for the \textit{local} Newton iteration}
\sectionfont{\small}
\label{sec:app_loctan}
For the iterative solution of the viscous evolution equation \eqref{eq:lokit-gls-3d} in the corrector step, the derivatives
\begin{equation}
 \diffp{\, r_\varrho}{\epse[\sigma]} = \delta_{\sigma \varrho} + \frac{\Delta t}{2 \, \etai}
 \, \diffp{\utaodev[\varrho]}{\epse[\sigma]}
 + \frac{\Delta t}{9 \etav} \, \diffp{\tr \utao}{\epse[\sigma]}
 \label{eq:locTan_allg}
\end{equation} 
with%
\footnote{For the implementation of the \textit{local} Newton iteration, no case-by-case analysis needs to be made accounting for whether there are multiple principal stretches and elastic principal stretches or not. Accordingly, the derivatives are given here with $N_\lambda=\nlame=3$ assumed. If algebraic multiplicities $\nu_\lambda,\nulame[\sigma]>1$ were explicitly considered, identical values for the derivatives would be obtained.}
\begin{align*}
 \diffp{\utaodev[\varrho]}{\epse[\sigma]} & = \sum_{p=1}^{\Nov} \muov \, \alov  \Bigg( \delta_{\varrho \sigma}  \, \left[ \lamed[\sigma] \right]^{\alov} - \frac{1}{3} \, \left[ \lamed[\varrho] \right]^{\alov} \\ 
 & - \frac{1}{3} \, \left[ \lamed[\sigma] \right]^{\alov} + \frac{1}{9} \, \sum_{\gamma=1}^3 \left[ \lamed[\gamma] \right]^{\alov} \Bigg)
 \comma
 \numberthis
 \label{eq:loktan_tauiso}
\end{align*}
\begin{equation}
 \diffp{\tr \utao}{\epse[\sigma]} = 3 \, \kov \, \jel^2
 \label{eq:loktan_tauvol}
\end{equation}
are required.
For the plane stress case, in addition, the derivatives
\begin{equation}
 \diffp{\, r_\varrho}{\varepsilon_3} = - \delta_{\varrho 3} \comma
\end{equation}
\begin{align*}
 \diffp{\prescript{0}{}{\uptau_3}}{\varepsilon_3} &=  
 \sum_{p=1}^{\Neq} \mueq \, \aleq  \Bigg( \frac{1}{3} \, \bar \lambda_3^{\aleq} +
 \frac{1}{9} \, \sum_{\gamma=1}^3 \bar \lambda_\gamma^{\aleq} \Bigg) 
 + \keq \, J^2
 \comma
 \numberthis
\end{align*}
\begin{equation}
 \diffp{\prescript{0}{}{\uptau_3}}{\epse[\sigma]}
 = \diffp{\utaodev[3]}{\epse[\sigma]}  + \frac{1}{3} \, \diffp{\tr \utao}{\epse[\sigma]}
\end{equation}
have to be evaluated.

\section{Material tangent}
\sectionfont{\small}
\label{sec:app_mattan}

The consistent Lagrangian material tangent
\begin{align*}
\tte C = 2 \, \diffp{\te T}{\te C} &=  g(d) \, \left(2 \, \diffp{\ute}{\te C} + 2 \, \diffp{\uto}{\te C}\right) \numberthis \\
& =: g(d) \, \left( \coeq + \coov \right)
\end{align*} 
is determined in order to enable the iterative solution of the weak form of balance of linear momentum~\eqref{eq:impb-s}.
In line with e.g. \cite{holzapfel2000}, the derivation of the tangent is performed assuming $N_\lambda=\nlame=3$ and the case of identical principal stretches or elastic principal stretches is then a posteriori addressed by means of L'Hôpital's rule.
%equilibrium part of the tangent
The equilibrium part of the virtually undamaged tangent is given by
\begin{align*}
 \coeq  = &  2 \, \diffp{\ute}{\te C} \\
  =  &
 \sum\limits_{\substack{\alpha \in \{1,2,3\} \\ \beta \in \{1,2,3\}}} \frac{1}{\lambda_\beta} \, \diffp{\prescript{0}{}{T}^\nv{eq}_\alpha}{\lambda_\beta} \, \ve N_\alpha \otimes \ve N_\alpha \otimes \ve N_\beta \otimes \ve N_\beta \\
 &+
 \sum\limits_{\substack{\alpha \in \{1,2,3\} \\ \beta \in \{1,2,3\} \setminus \alpha }}
 \frac{\prescript{0}{}{T}^\nv{eq}_\beta-\prescript{0}{}{T}^\nv{eq}_\alpha}{\lambda_\beta^2-\lambda_\alpha^2} 
 \, \ve N_\alpha \otimes \ve N_\beta  \\ 
 & \qquad \qquad   \otimes \left( \ve N_\alpha \otimes \ve N_\beta + \ve N_\beta \otimes \ve N_\alpha \right)
 %\comma
 \numberthis
 \label{eq:eqTan-allg}
\end{align*} 
wherein $\ve N_\varrho$ denote the orthonormal eigenvectors of $\te{C}$,%
\footnote{It is assumed that an appropriate orthonormalization method is used in case of multiple principal stretches.}
see e.g. \cite{miehe1993,miehe1998} or \cite{kalina2020} for a derivation of the derivatives of principle stretches and projection tensors.
Into this expression \eqref{eq:eqTan-allg}, for the specific model under consideration,
\begin{align*}
 \frac{1}{\lambda_\beta} &\,\diffp{\prescript{0}{}{T}^\nv{eq}_\alpha}{\lambda_\beta} 
 =
 - \frac{2}{\lambda_\alpha^2} \, \prescript{0}{}{T}^\nv{eq}_\alpha \, \delta_{\alpha \beta}
 + \frac{1}{\lambda_\alpha^2 \, \lambda_\beta^2} \, \Biggl[ J^2 \, \keq \\
 & + \sum_{p=1}^{\Neq} \mueq \, \aleq \left(
    \bar \lambda_\alpha^{\aleq} \, \delta_{\alpha \beta}
    -\frac{1}{3} \bar \lambda_\alpha^{\aleq}
    -\frac{1}{3} \bar \lambda_\beta^{\aleq}
    +\frac{1}{9} \sum_{\varrho=1}^{3}
    \bar\lambda_\varrho^{\aleq} \right)
\Biggl]
\numberthis
\end{align*} 
can be inserted.
In case of multiple principal stretches, i.e. $\exists \beta \neq \alpha : \lambda_\beta = \lambda_\alpha$, the second term in \eqref{eq:eqTan-allg} can be evaluated making use of L'Hôpital's rule \cite{holzapfel2000}
\begin{align*}
 \lim_{\lambda_\beta \rightarrow \lambda_\alpha} = \frac{\prescript{0}{}{T}^\nv{eq}_\beta-\prescript{0}{}{T}^\nv{eq}_\alpha}{\lambda_\beta^2-\lambda_\alpha^2}
 &= \lim_{\lambda_\beta \rightarrow \lambda_\alpha} \frac{1}{2 \, \lambda_\beta} \left( \diffp{\prescript{0}{}{T}^\nv{eq}_\beta}{\lambda_\beta} - \diffp{\prescript{0}{}{T}^\nv{eq}_\alpha}{\lambda_\beta}  \right) \\
 & = - \frac{1}{\lambda_\alpha^2} \, \prescript{0}{}{T}^\nv{eq}_\beta + \frac{1}{2 \, \lambda_\alpha^4} 
 \, \sum_{p=1}^{\Neq} \mueq \, \aleq \, \bar \lambda_\alpha^{\aleq}
 \point
 \numberthis
\end{align*}
% -- ov part of tangent --
Following \cite{reese1998}, for the derivation of $\coov$, a virtually undamaged over-stress tensor
\begin{equation}
 \utoh = \fvio \cdot \uto \cdot \fvio^\top
\end{equation}
is introduced with reference to the intermediate configuration defined by the viscous deformation gradient of the previous time step $\fvio$, i.e. based on the decomposition of the deformation gradient at increment $n$ into
\begin{equation}
 \prescript{}{n}{\te F} = \prescript{}{n}{\fetr} \cdot \fvio  \point
\end{equation} 
With
\begin{equation}
 \celtrh = \fetr^\top \cdot \fetr \comma
\end{equation}
the over-stress part of the virtually undamaged material tangent then can be written as
\begin{align*}
 \coov_{KLMN} = 2 & \fvioab{K\gamma}^{-1} \, \fvioab{M \alpha}^{-1} \\
 & \cdot \underbrace{\diffp{\utohab{\gamma \delta}}{\celtrhab{\, \alpha \beta}}}_{=: \coovh_{\gamma \delta \alpha \beta}/2} 
  \,  \fvioab{L \delta}^{-1} \, \fvioab{N \beta}^{-1} \comma
 \numberthis
\end{align*}
wherein the Einstein summation convention applies for double indices.
From this, the over-stress tangent in terms of the intermediate configuration described by $\fvio$ can be defined to
\begin{equation}
 \coovh := \diffp{\utoh}{\celtrh} \point
\end{equation} 
In analogy to \eqref{eq:eqTan-allg}, this contribution to the material tangent is given by
\begin{align*}
 \coovh  & = 
  \sum\limits_{\substack{\alpha \in \{1,2,3\} \\ \beta \in \{1,2,3\}}} 
  \frac{1}{\lametr[\beta]} \, \diffp{\prescript{0}{}{\breve T}^\nv{ov}_\alpha}{\lametr[\beta]} 
  \, \Neh[\alpha] \otimes \Neh[\alpha] \otimes \Neh[\beta] \otimes \Neh[\beta] \\
 &+
 \sum\limits_{\substack{\alpha \in \{1,2,3\} \\ \beta \in \{1,2,3\} \setminus \alpha }}
 \frac{\prescript{0}{}{\breve T}^\nv{ov}_\beta-\prescript{0}{}{\breve T}^\nv{ov}_\alpha}{\left(\lametr[\beta]\right)^2-\left(\lametr[\alpha]\right)^2} \, \\
 & \hspace{2cm} \Neh[\alpha] \otimes \Neh[\beta] \otimes \left( \Neh[\alpha] \otimes \Neh[\beta] + \Neh[\beta] \otimes \Neh[\alpha] \right)
 \comma
 \numberthis
 \label{eq:ovTanh-allg}
\end{align*}
wherein $\Neh[\varrho]$ denote the orthonormal eigenvectors of $\celtrh$ and $\prescript{0}{}{\breve T}^\nv{ov}_\sigma$ are the eigenvalues of $\utoh$.
The first term in \eqref{eq:ovTanh-allg} can be rewritten making use of
\begin{equation}
  \frac{1}{\lametr[\beta]} \, \diffp{\prescript{0}{}{\breve T}^\nv{ov}_\alpha}{\lametr[\beta]}
  =
  - \frac{2}{\left(\lametr[\alpha]\right)^4} \, \utaoj[\alpha] \, \delta_{\alpha \beta} 
  + \frac{1}{\left(\lametr[\alpha] \, \lametr[\beta]\right)^2} \diffp{\utaoj[\alpha]}{\epsetr[\beta]}
  \point
  \label{eq:deriv_trialstr}
\end{equation} 
Furthermore, for the derivatives with respect to the trial stretch quantities, use of
\begin{equation}
 0 = \diffp{r_\varrho}{\epsetr[\sigma]} \comma
\end{equation}
which holds if the \textit{local} Newton iteration has converged towards zero, is made.
This assumption leads to
\begin{equation}
 \diffp{\epse[\varrho]}{\epsetr[\sigma]} = \left(\diffp{r_\sigma}{\epse[\sigma]} \right)^{-1} \comma
\end{equation} 
which is given by \eqref{eq:locTan_allg} and further specified in Appendix~\ref{sec:app_loctan}. Accordingly, the derivative $\partial{\utaoj[\alpha]}/\partial {\epsetr[\beta]}$ in \eqref{eq:deriv_trialstr} can be expressed as
\begin{equation}
 \diffp{\utaoj[\alpha]}{\epsetr[\beta]} = \sum_{\varrho=1}^3 \diffp{\utaoj[\alpha]}{\epse[\varrho]} \, \left(\diffp{r_\varrho}{\epse[\beta]} \right)^{-1}
\end{equation}
with
\begin{equation}
 \diffp{\utaoj[\alpha]}{\epse[\varrho]} 
 =
 \diffp{\utaodev[\alpha]}{\epse[\varrho]}  + \frac{1}{3} \, \diffp{\tr \utao}{\epse[\varrho]}
 \label{eq:abl-tauov_alpha}
\end{equation}
and the two contributions in \eqref{eq:abl-tauov_alpha} specified by \eqref{eq:loktan_tauiso} and \eqref{eq:loktan_tauvol}.
% multiple elastic principal stretches
In case of multiple elastic principal stretches, i.e. $\exists \beta \neq \alpha : \lame[\beta] = \lame[\alpha]$, for the treatment of the second term in \eqref{eq:ovTanh-allg}, the same procedure applies as outlined above for the case of multiple principal stretches.
In particular, L'Hôpital's rule reads
\begin{equation}
  \lim_{\lametr[\beta] \rightarrow \lametr[\alpha]} \frac{\prescript{0}{}{\breve T}^\nv{ov}_\beta-\prescript{0}{}{\breve T}^\nv{ov}_\alpha}{\left(\lametr[\beta]\right)^2-\left(\lametr[\alpha]\right)^2}
  =
  \lim_{\lametr[\beta] \rightarrow \lametr[\alpha]} \frac{1}{2 \, \lametr[\beta]} \left( \diffp{\prescript{0}{}{\breve T}^\nv{ov}_\beta}{\lametr[\beta]} - \diffp{\prescript{0}{}{\breve T}^\nv{ov}_\alpha}{\lametr[\beta]}  \right)
\end{equation}
with the respective derivatives given by \eqref{eq:deriv_trialstr}.
\end{document}